key words:  CGL equation, random perturbation, averaging, effective equations, stationary measure, inviscid limit
\documentclass[12pt,draft]{article}
\usepackage{amsmath,amsfonts,amsthm,amssymb,amscd,colordvi}
\renewcommand{\Re}{\mathop{\rm Re}\nolimits}  

\newcommand{\p}{\partial}

\newcommand{\e}{\varepsilon}
\newcommand{\vk}{\varkappa}

\newcommand{\vp}{\varphi}
\newcommand{\gr}{\gamma_R}
\newcommand{\gi}{\gamma_I}
\newcommand{\la}{\lambda}

\newcommand{\bb}{\mbox{\boldmath$\beta$}}
\newcommand{\R}{{\mathbb R}}
\newcommand{\C}{{\mathbb C}}
\newcommand{\IP}{{\bf P}}
\newcommand{\Z}{{\mathbb Z}}
\newcommand{\E}{{\bf E}}
\newcommand{\cE}{{\cal E}}
\newcommand{\T}{{\mathbb T}}
\newcommand{\N}{{\mathbb N}}

\newcommand{\hA}{\widehat A}
\newcommand{\OR}{\Omega_R}

\newcommand{\cD}{{\cal D}}
\newcommand{\cF}{{\cal F}}

\newcommand{\cH}{{\cal H}}
\newcommand{\cL}{{\cal L}}

\newcommand{\strela}{\rightharpoonup}

\newcommand{\as}{\quad\mathop{\rm as} \nolimits\quad}
\newcommand{\diag}{\mathop{\rm diag}\nolimits}

\def\12{\tfrac12}

\def\lan{\langle}
\def\ran{\rangle}

\theoremstyle{plain}
\newtheorem{theorem}{Theorem}[section]
\newtheorem{lemma}[theorem]{Lemma}
\newtheorem{proposition}[theorem]{Proposition}

\theoremstyle{definition}

\theoremstyle{remark}

\newtheorem{example}[theorem]{Example}

\numberwithin{equation}{section}

\begin{document}
\author{Sergei B. Kuksin}
\title{Weakly nonlinear stochastic CGL equations.}
\date{}
\maketitle

\hfill{\it Dedicated to Claude Bardos on his 70-th
birthday } \medskip

\begin{abstract}
We consider the linear Schr\"odinger equation under periodic boundary condition, driven by a random 
force and damped by a quasilinear damping:
$$
\frac{d}{dt}u+i\big(-\Delta+V(x)\big)
u=\nu  \Big( \Delta u-\gr  |u|^{2p}u-i\gi |u|^{2q}u \Big) 
+\sqrt\nu\, \eta(t,x).\qquad (*)
$$
The force $\eta$ is white in time and smooth in $x$. We are concerned with the limiting, as $\nu\to0$, 
behaviour of its solutions on long time-intervals $0\le t\le\nu^{-1}T$, and with behaviour of these solutions 
under the double limit $t\to\infty$ and $\nu\to0$. We show that these two limiting behaviours may be 
described  in terms of solutions for the
{\it  system of effective equations for $(*)$ } which is a well posed 
semilinear stochastic heat equation with a non-local nonlinearity
and a smooth additive noise, written in  Fourier coefficients. The effective equations do not depend on the 
Hamiltonian part of the perturbation $-i\gi|u|^{2q}u$ (but depend on the dissipative part $-\gr|u|^{2p}u$). 
If $p$ is an integer, they may be written explicitly. 

\end{abstract}

\tableofcontents

\section{Introduction}\label{s0}
In \cite{KP08, K10} we have considered the KdV equation on a circle, perturbed by a random force and a viscous 
damping. There we suggested auxiliary {\it effective equations} which are well posed and describe long-time 
behaviour of solutions for the perturbed KdV  through a kind of averaging.

In this work we apply the method  of \cite{KP08,K10} to a weakly nonlinear situation when the unperturbed 
equation is not an integrable nonlinear PDE (e.g.  KdV),  but a linear Hamiltonian PDE with a generic spectrum.
Since analytic properties of the latter are easier and better understood then those of the former, in the 
weakly nonlinear situation we understand better properties of the effective system and its relation with the 
original equation. Accordingly  we can go further in analysis of long time behaviour of solutions.

More precisely, we are concerned with $\nu$-small dissipative stochastic perturbations of the 
space-periodic linear Schr\"odinger equation
\begin{equation}\label{0.1}
\frac{d}{dt}u+i(-\Delta u+V(x))u=0,\qquad x\in\T^d,
\end{equation}
i.e. with equations
\begin{equation}\label{0.3}
\begin{split}
\frac{d}{dt}u+iAu=\nu  \Big( \Delta u-\gr f_p(|u|^2)u-i\gi f_q(|u|^2)u \Big) 
+\sqrt\nu\, \eta(t,x),\quad x\in\T^d,  
\end{split}
\end{equation}
where $\eta(t,x)= \frac{d}{dt}\sum_{j=1}^\infty b_j\bb_j(t)e_j(x)$.
Here $Au=A_Vu=-\Delta u+V(x)u$ and  the potential $V(x)\ge1$ is sufficiently smooth; the real numbers
$p,q$ are non-negative, the functions $f_p(r)$ and $f_q(r)$ are the monomials $|r|^p$ and $|r|^q$, 
smoothed out near zero, and the constants $\gr, \gi$ satisfy 
\begin{equation}\label{1.0}
\gr, \gi\ge0,\qquad \gr+\gi=1.
\end{equation}
If $\gr=0$, then due to the usual difficulty with the zero-mode of the solution $u$, the term $\Delta u$ 
in the r.h.s. should be modified to $\Delta-u$. The functions $\{e_j(x),\ j\ge1\}$ in the definition of the random force 
form 
the real trigonometric base of $L_2(\T^d)$, the real numbers $b_j$ decay sufficiently fast  to zero when
$j$ grows, and $\{\bb_j(t),\ j\ge1\}$, are the standard complex Wiener processes. So the noise $\eta$
 is white in time and sufficiently smooth in $x$. It is convenient to pass to the slow time
$\tau=\nu t$ and write the equation as 
\begin{equation}\label{0.4}
\begin{split}
\dot u+\nu^{-1} iAu= \Delta u-\gr f_p(|u|^2)u-i\gi f_q(|u|^2)u
+  \eta(\tau,x) ,
\end{split}
\end{equation}
where $\dot u=du/d\tau$. The equation is supplemented with the initial condition
\begin{equation}\label{0.5}
u(0,x)=u_0(x). 
\end{equation}
It is known that under certain restrictions on $p,q$ and $d$ the problem \eqref{0.4}, \eqref{0.5} has a 
unique solution $u^\nu(\tau,x)$, $\tau\ge0$, and  eq.~\eqref{0.4} has a unique stationary measure $\mu^\nu$.
We review these results in Section~\ref{s1} (there attention is given to the 1d case, while higher-dimensional
equations are only briefly discussed).

Let $\{\vp_k,\ k\ge1\}$, and  $\{\lambda_k,\ k\ge1\}$, be the eigenfunctions and eigenvalues of $A_V$,
$1\le\lambda_1\le\lambda_2\le\dots$.  We say that a potential $V$ is {\it nonresonant} if 
$\ 
\sum_{j=1}^\infty \lambda_js_j\ne0
$
for every finite non-zero  integer vector $(s_1, s_2,\dots)$. In Sections~\ref{s3.1}, \ref{s3.2} we 
show that nonresonant potentials are typical both in the sense of Baire and in the sense of measure. 
Assuming that $V$ is nonresonant we are interested in two questions:
\medskip

\noindent
Q1. What is the limiting behaviour as $\nu\to0$ of solutions $u^\nu(\tau,x)$ on long time-intervals
$0\le\tau\le T$?
\smallskip

\noindent 
Q2. What is the limiting behaviour of the stationary measure $\mu^\nu$ as $\nu\to0$?
\medskip

For any complex function $u(x),\, x\in\T^d$, denote by 
$
\ \Psi(u)=v=(v_1,v_2,\dots)
$
the complex vector of its Fourier coefficients with respect to the basis $\{\vp_k\}$, i.e. 
$u(x)=\sum v_j\vp_j$. Denote
\begin{equation}\label{ac-an}
I_j=\frac12|v_j|^2,\quad \vp_j=\text{Arg}\,v_j, \qquad j\ge1. 
\end{equation}
Then $(I,\vp)\in\R^\infty_+\times \T^\infty$ are the action-angles for the linear  equation \eqref{0.1}.
The $v$-  and $(I,\vp)$-variables are convenient to study the two questions above. Writing \eqref{0.4}
in the $(I,\vp)$-variables we arrive at the following system:
\begin{equation}\label{0.7}
\begin{split}
\frac{d}{d\tau}I_j=\dots,\qquad \frac{d}{d\tau}\vp_j=  \lambda_j+ \dots,
\end{split}
\end{equation}
where the dots stand for terms of order one (stochastic and deterministic).  
We have got slow/fast stochastic equations to which the principle of averaging 
 is formally applicable (e.g., see \cite{AKN, LochM} for the classical deterministic averaging and 
 \cite{Khas68,FW98} for the stochastic averaging). Denoting $I^\nu_j(\tau)=I_j(u^\nu(\tau))$ and 
 averaging in $\vp$  the $I$-equations in \eqref{0.7}, using the rules of the stochastic calculus  \cite{Khas68,FW98}
 and  following the arguments in \cite{KP08}, we show in Section~\ref{s5} that along sequences 
$\nu_j\to0$ we have the convergences 
\begin{equation}\label{0.8}
\cD(I^{\nu_j}(\cdot))\strela\cD(I^0(\cdot)),
\end{equation}
where the limiting process $I^0(\tau)$,  $0\le\tau\le T$,
is a weak solution of the averaged $I$-equations.  As in the KdV-case
the averaged  equations are singular and we do not know if their solution is unique. So we do not know if the convergence 
\eqref{0.8} holds as $\nu\to0$.    To continue  the analysis we
write eq.~\eqref{0.4} in the $v$-variables
\begin{equation}\label{0.9}
\dot v_k+i\nu^{-1}\lambda_k v_k=P_k(v)+\sum_{j\ge1}B_{kj}\dot\bb_j(\tau),
\end{equation}
where the drift $P_k$ and the dispersion $B_{kj}$ are written explicitly in terms of the r.h.s. of eq.~\eqref{0.4}.
It turns out that the Hamiltonian term $-i\gi f_q(|u|^2)u$ contributes to $P(v)$ a term which 
disappears in the averaged $I$-equations. We remove it from $P(v)$ and denote the rest $\tilde P(v)$. 
For any vector $\theta=(\theta_1,\theta_2,\dots)\in\T^\infty$ denote by $\Phi_\theta$ the linear 
transformation of the space of complex vectors $v$ which multiplies each component $v_j$ by
$e^{i\theta_j}$.  Following \cite{K10} we average the vector field $\tilde P$  by actions of the transformations 
$\Phi_\theta$ and get the {\it effective drift}  
$\ 
R(v)=\int_{\T^\infty}\Phi_{-\theta}\tilde P(\Phi_\theta v)\,d\theta.
$
In Section \ref{s7.1} we show that 
\begin{equation}\label{0.10}
R_k(v)=-\lambda_k v_k+R^0_k(v),
\end{equation}
where $R^0(v)$ is a smooth locally Lipschitz nonlinearity. 

Since the noise in \eqref{0.9} is additive (i.e., the matrix $B$ is $v$-independent), then construction of the 
effective dispersion, given in \cite{K10} for non-additive noise, simplifies significantly and defines the 
effective noise for eq.~\eqref{0.9}   whose  $k$-th component  equals 
$\ 
\left(\sum_l b_l^2\Psi^2_{kl}\right)^{1/2} d\bb_k(\tau).
$ 
Accordingly the effective equations for \eqref{0.4} become
\begin{equation}\label{0.11}
\dot v_k=R_k(v)\,d\tau+\Big(\sum_l b_l^2\Psi^2_{kl}\Big)^{1/2} d\bb_k(\tau),\qquad k\ge1. 
\end{equation}
By construction this system is invariant under rotations: if $v(\tau)$ is its weak solution, 
then $\Phi_\theta v(\tau)$ also is a weak solution. Due to \eqref{0.10} this is the heat equation
$\  \dot u=-Au$ for a complex function $u(\tau,x)$, perturbed by a non-local smooth nonlinearity and a 
non-degenerate smooth noise, written in terms of the complex Fourier coefficients $v_j$.  It turns out 
to be a monotone equation, so  its solution is unique (see Section~\ref{s7.2}). 

In particular, if in \eqref{0.4}  $p=1$, then the system of effective equations takes the form 
\begin{equation}\label{0.12}
\dot v_k= -v_k \big((\lambda_k-M_k)+\gr\sum|v_l|^2L_{kl}\big)\,d\tau
+\Big(\sum_l b_l^2\Psi^2_{kl}\Big)^{1/2} d\bb_k(\tau),\quad k\ge1,
\end{equation}
where $M_k=\int V(x)\vp^2(x)\,dx$ and 
$\ L_{kl}=(2-\delta_{kl})\int \vp_k^2(x)\vp_l^2(x)\,dx$.  See Example~\ref{e7.1}  (the 
calculations, made there for $d=1$, remain the same for $d\ge2$). 

It follows directly from the construction of effective equations that actions 
$\{I(v_k(\tau)) = \tfrac12|v_k(\tau)|^2 ,\ k\ge1\}$ 
of any solution $v(\tau)$ of \eqref{0.11} is a solution of the system of averaged $I$-equations. On the 
contrary, every solution $I^0(\tau)$ of the averaged $I$-equations, obtained as a limit \eqref{0.8}, 
can be lifted to a weak solution of \eqref{0.11}. Using the uniqueness we get 

\begin{theorem}\label{tA}
Let $I^\nu(\tau)=I(u^\nu(\tau))$, where $u^\nu(\tau)$, $0\le\tau\le T$, is a solution of \eqref{0.4}, \eqref{0.5}. Then 
$\lim_{\nu\to0}\cD(I^\nu(\cdot))= \cD(I^0(\cdot))$, where $I^0(\tau)$, $0\le\tau\le T$, is a weak solution of the 
averaged $I$-equations. Moreover, there exists a unique solution $v(\tau)$ of \eqref{0.11}  such 
that  $v(0)=v_0=\Psi(u_0)$ and $\cD(I(v(\cdot))=\cD(I^0(\cdot))$, where $I(v(\tau))_j=\tfrac12|v_j(\tau)|^2$.
\end{theorem}

The solutions $I^0(\tau)$ and $v(\tau)$ satisfy some apriori estimates, see Theorem~\ref{t7.3}. 
Concerning distribution of the angles $\vp(u^\nu(\tau))$ and their joint distribution with the actions see
Section~\ref{s5.4}. 
\medskip

Now let $\mu^\nu$ be the unique stationary measure for eq.~\eqref{0.4} and ${u'}^\nu$ be a corresponding stationary
solution, $\cD({u'}^\nu(\tau))\equiv\mu^\nu$. As above, along sequences $\nu_j\to0$ the actions 
${I'}^{\nu_j}(\tau)=I({u'}^{\nu_j}(\tau))$ converge in distribution to  stationary solutions $I'(\tau)$ of the averaged 
$I$-equations.  These solutions can be lifted to  stationary weak solutions $v'(\tau)$ of effective equations 
\eqref{0.11}.  Since that system is monotone, then its stationary measure $m$ is unique. So the limit above holds
as $\nu\to0$.  As the effective system is rotation invariant, then in the $(I,\vp)$-variables its unique stationary
measure has the form $dm=m_I(dI)\times d\vp$. It turns out that the measure $\lim_{\nu\to0}\mu^\nu$ also 
has the rotation-invariant form and we arrive at the following result (see Theorem~\ref{t6.3}, \ref{t6.4} for a 
precise statement):

\begin{theorem}\label{tB}
When $\nu\to0$ we have the convergences
$\ 
\cD\left( I({u'}^\nu(\cdot)\right)\strela \cD I(v'(\cdot))$ 
and
$
\Psi\circ\mu^\nu\strela m$, where  $dm=m_I(dI)\times d\vp$. 
\end{theorem}

Accordingly  every solution $u^\nu(\tau)$ of \eqref{0.3} obeys the following double limit
\begin{equation}\label{0.20}
\lim_{\nu\to0}\lim_{t\to\infty}\cD(u^\nu(t))= \Psi^{-1}  \circ m.
\end{equation}

By  Theorems \ref{tA} and \ref{tB}, the actions $I(u^\nu(\tau))$ of a solution $u^\nu$ of 
\eqref{0.4}, \eqref{0.5} converge in distribution to those of a solution $v(\tau)$ of the effective system
\eqref{0.11} with $v(0)=\Psi(u_0)$, both for $0\le\tau\le T$ and when $\tau\to\infty$.  We conjecture that 
this convergence hold for each $\tau\ge0$, uniformly in $\tau$ (the space of measures is equipped 
with  the Wasserstein distance). 

In Example \ref{e6.4} we discuss Theorem~\ref{tB} for equations with $p=1$, when the effective 
equations become \eqref{0.12}. In  particular, we show that Theorem~\ref{tB} implies that in equations
\eqref{0.3} with small $\nu$ there is no direct or inverse cascade of energy.

In Example \ref{e6.3} we discuss Theorem \ref{tB} for the case $\gr=0$ (when the nonlinear part of the 
perturbation is Hamiltonian)  and its relation to the theory of weak turbulence.

We note that the effective equations \eqref{0.11} depend on the potential $V(x)$ in a regular way
and are well defined without assuming that $V(x)$ is nonresonant (cf. equations \eqref{0.12}).
In particular, if $V^M(x)\to1$ as $M\to\infty$, where each $V^M(x)\ge1$  is a non-resonant potential,
then in \eqref{0.20} $m^M\strela m(1)$, where $m(1)$ is a unique stationary measure for eq.~\eqref{0.12}
with $V(x)\equiv1$. In  this equation $\Psi_{kl}=\delta_{k,l}$, $M_k\equiv1$ and the constants $L_{kl}$ 
can be  written down explicitly. 
\smallskip

In Section \ref{s8} we show that Theorems \ref{tA}, \ref{tB} remain true for 1d equations with 
non-viscous damping (when $\Delta u$ in the l.h.s. of \eqref{0.3} is removed, but $\gr>0$). 
\medskip

\noindent
{\bf Inviscid Limit.}  A stationary measure $\mu^\nu$  for eq. \eqref{0.4} also is  stationary for the 
fast-time equation \eqref{0.3}. Let  $U^\nu(t)$ be a corresponding stationary solution, 
$\cD U^\nu(t)\equiv\mu^\nu$. It is not hard to see that the system of solutions $U^\nu(t)$ is tight on any 
finite time-interval $[0, \tilde T]$. Let $\{U^{\nu_j}, \nu_j\to0\}$, be a converging subsequence, i.e. 
$$
\cD(U^{\nu_j}) \strela Q^*,\qquad \mu^{\nu_j}\strela \mu^*.
$$
Then $\mu^*$ is an invariant measure for the linear equation \eqref{0.1} and $Q^*=\cD(U^*(\cdot))$, where 
$U^*(t), 0\le t\le  \tilde T$,  is a stationary process such that $\cD(U^*(t))\equiv\mu^*$ and every trajectory of $U^*$
is a solution of \eqref{0.1}. The limit $\cD(U^{\nu_j} )\strela \cD(U^*)$ is the {\it inviscid limit for eq.~\eqref{0.3}}. 
Eq.~\eqref{0.1} has plenty of invariant measures: if we write it in the action-angle variables \eqref{ac-an}, 
then every measure of the form $m(dI)\times d\vp$ is invariant (see \cite{KS04J} for the more complicated  inviscid limit for 
nonlinear Schr\"odinger equation).  Theorem~\ref{tB} explains which one is 
chosen by eq.~\eqref{0.3} for the limit $\lim_{\nu\to0}\mu^\nu$. 

The inviscid limit for the damped/driven KdV equation, studied in \cite{KP08, K10} is similar: the limit of the
stationary measures for the perturbed equations is a stationary measure of the corresponding effective equations. 
Due to a complicated structure of the nonlinear Fourier transform which integrates KdV,
uniqueness of their invariant measure is not proved yet. So  the
final results concerning the damped/driven KdV are less complete than those for the weakly perturbed CGL
equation in Theorem~\ref{tB}.

Finally consider the damped/driven 2d Navier-Stokes equations with  a small viscosity $\nu$ and a 
random force, similar to the forces above and proportional to $\sqrt\nu$:
\begin{equation}     \label{NSE}
\begin{split}
 v'_t - \nu \Delta v+(v\cdot\nabla)v+\nabla  p=\sqrt\nu\,
\eta(t,x);\qquad 
\text{div}\, v=0, \; v\in\R^2,\;
\; x\in \T^2
 \end{split}
\end{equation}
 It is known that \eqref{NSE} has a unique stationary measure $\mu^\nu$, the family of measures
 $\{\mu^\nu,0<\nu\le1\}$ is tight, and every limiting measure $\lim_{\nu_j\to0}\mu^{\nu_j}$ is a 
 non-trivial invariant measure for the 2d Euler equation  \eqref{NSE}${}_{\nu=0}$,
  see Section~5.2 of \cite{KS}. Hovewer it is non-clear if the 
 limiting measure is unique and how to single it out among all invariant measures of the  Euler
 equation. The research \cite{KP08,K10} was motivated by the belief that the damped/driven KdV is a 
 model for \eqref{NSE}. Unfortunately, we still do not know up to what extend the 
 description of the inviscid limit for the damped/driven KdV and for weakly nonlinear CGL in terms
 of the effective equations is relevant for the inviscid limit of the 2d hydrodynamics. 

\medskip

\bigskip
\noindent {\it Agreements.} Analyticity of maps $B_1\to B_2$ between
Banach spaces $B_1$ and $B_2$, which are the real parts of complex
spaces $ B_1^c$ and $B_2^c$, is understood in the sense of
Fr\'echet. All analytic maps which we consider possess the following
additional property: for any $R$ a map analytically extends to a
complex $(\delta_R>0)$--neighbourhood of the ball $\{|u|_{B_1}<R\}$
in $B_1^c$.   

\noindent {\it  Notations.}  $\chi_A$ stands for the indicator
function of a set $A$ (equal 1 in $A$ and equal 0 outside $A$). By
$\vk(t)$ we denote various functions of $t$ such that $\vk(t)\to0$
when $t\to\infty$, and by $\vk_\infty(t)$ denote functions $\vk(t)$
such that $\vk(t)=o(t^{-N})$ for each $N$. We write
$\vk(t)=\vk(t;R)$ to indicate that $\vk(t)$ depends on a parameter
$R$.

\medskip\par
\noindent{\it Acknowledgments.} I wish to thank for discussions and advice Patrick Gerard, 
Sergey Nazarenko, Andrey Piatnitski and Vladimir Zeitlin. 
This work was supported by  l'Agence Nationale de la Recherche through the grant  ANR-10-BLAN 0102.

\section{Preliminaries}\label{s1}
\subsection{Apriori  estimates.}\label{s1.1}
We consider the 1d \ CGL equation on a segment $[0,\pi]$ with  a conservative 
linear part of order one and a 
 small nonlinearity. The equation is supplemented with Dirichlet boundary conditions which we interpret 
as odd $2\pi$-periodic boundary conditions. 
 Introducing the 
slow time $\tau=\nu t$ (cf. Introduction) we write the equation as follows:
\begin{equation}\label{1.1}
\begin{split}
\dot u+i\nu^{-1}\big(-u_{xx}&+V(x)u\big)=\vk u_{xx}-\gr|u|^{2p}u-i\gi|u|^{2q}u
\\
&+\frac{d}{d\tau}\sum_{i=1}^\infty b_j\bb_j(\tau)e_j(x),\qquad
u(x)\equiv u(x+2\pi)\equiv -u(-x).
\end{split}
\end{equation}
Here $\dot u=\frac{d}{d\tau}u$, $\ p,q\in\Z_+:=\N \cup\{0\}$ 
(only for simplicity, see next section),  $\vk>0$, 
constants $\gr$ and $\gi$ satisfy \eqref{1.0} and 
$\R\ni V(x)\ge0$ is a sufficiently smooth even  $2\pi$-periodic function, $\{e_j, j\ge1\}$ is the sine-basis, 
$$
e_j(x)=\frac1{\sqrt\pi}\sin jx,
$$
and $\bb_j, j\ge1$, are  standard independent complex Wiener processes. That is, $\bb_j(\tau)=\beta_{j}(\tau)+i\beta_{-j}(\tau)$,
where $\beta_{\pm j}(\tau)$ are standard independent  real Wiener processes. 
Finally, the real numbers $b_j$ all are non-zero and decay when $j$ grows in such a way that $B_1 <\infty$, where
$$
B_r:=2\sum_{j=1}^\infty j^{2r} b_j^2\leq\infty \quad\text{for}\quad r\geq 0.
$$
By $\cH^r$, $r\in \R$ we denote the Sobolev space of order $r$ of complex odd periodic functions and 
provide it 
with the homogeneous norm $\|\cdot\|_r$,  
$$
\|u\|^2_r  
= \sum^\infty_{l=1} |u_l|^2 l^{2r}, \quad \|\cdot\|_0 =\|\cdot\| 
$$
(if $r\in\N$, then $\|u\|_r = \left|\frac{\p^r u}{\p x^r}\right |_{L_2}$).

Let $u(t, x)$ be a solution of \eqref{1.1} such that
$\ 
u(0, x)=u_0.  
$
Applying Ito's formula to $\frac 12 \|u\|^2$ we get that
\begin{equation}\label{1.2}
\begin{split}
d\left(\frac 12 \|u\|^2\right)= (-\gamma_r |u|^{2p+2}_{2p+2} -\vk \|u\|_1^2+\frac12 B_0)d\tau + d M(\tau),
\end{split}
\end{equation}
where $M(\tau)$ is the martingale $\int^\tau _0\sum b_j u_j\cdot d\bb_j (\tau).$ 
Here $|u|_r$  stands for 
the $L_r$-norm, $1\leq r\leq \infty$, and for complex numbers $z_1$, $z_2$ we denote by
$z_1 \cdot z_2 $ their real scalar product, 
$$
z_1 \cdot z_2 = \Re z_1\overline {z_2}.
$$
So $(u_{j} + iu_{-j})\cdot(d\beta_{j}+id\beta_{-j})=u_{j}d\beta_{j} + u_{-j} d\beta_{-j}$. From \eqref{1.2} we get in
the usual way (e.g., see Section~2.2.3  in \cite{KS}) that 
\begin{equation}\label{1.3}
\E   e^{\rho_\vk\|u(\tau)\|^2} \leq C(\vk,B_0,\|u_0\|) \quad \forall \tau\geq 0
\end{equation}
for a suitable $\rho_\vk>0$, uniformly in $\nu>0$.

Denoting
$$
\cE (\tau)=\frac12\|u(\tau)\|^2+\gamma_r\int^\tau_0 |u|_{2p+p}^{2p+p} ds +\frac{\vk}2\int_0^\tau \|u\|_1^2 ds
$$
and noting that the characteristic of the martingale $M$ is
$\ 
\langle M \rangle (\tau)=\sum b_j^2 |u_j|^2\leq b_M^2\|u\|^2$, where $b_M=\max |b_j|$, 
we get from \eqref{1.2} that
\begin{equation*}
\begin{split}
\cE (\tau)&\leq \tfrac12\|u_0\|^2 + \tfrac12B_0\tau +M(\tau)-  \frac{\vk}2\int^\tau_0 \|u\|^2\, ds\\
&\le \tfrac12 \|u_0\|^2 +\tfrac12B_0\tau +\vk^{-1} b^2_M \left[(\vk b_M^{-2} M(\tau))-\frac12 \langle\vk b_M^{-2} M\rangle (\tau)\right].
\end{split}
\end{equation*}
Applying in a standard way the exponential supermartingale estimate to the term in the square bracket in the r.h.s.
(e.g., see  \cite{KS}, Section~2.2.3 ), we get that
\begin{equation}\label{1.5}
{\IP}\{
\sup_{\tau\ge0} (\cE (\tau) -\tfrac12 B_0\tau) \geq \tfrac12\|u_0\|^2 +\rho \}
\leq e^{-2\vk \rho b_M^{-2}},
\end{equation}
for any $\rho >0$.

Now let us re-write eq. \eqref{1.1} as follows:
\begin{equation}\label{1.6}
\dot u + i \nu^{-1}\big (-u_{xx}+V(x) u +\nu \gamma_I |u|^{2q} u\big)=\vk u_{xx} - \gamma_R |u|^{2p} u +
\frac{d}{d\tau} \sum b_j \beta_j (\tau)e_j.
\end{equation}
The l.h.s. is a Hamiltonian system with the hamiltonian $-\nu^{-1} H(u)$,
$$
H(u)=\frac12\langle Au, u\rangle +\gamma_I \frac{\nu}{2q+2}\int |u|^{2q+2} dx,\quad A=-\frac{\partial^2}{\partial x^2} + V(x).
$$
For any $ j\in \N$ we denote
$$
{\|u\|_r'}^2 = \langle A^r u, u\rangle.
$$
Then $dH(u)(v)=\langle A u, v\rangle +\gamma_I \nu \langle |u|^{2q} u, v\rangle$ and 
$$
\tfrac12\cdot 2\sum_{j=1}^\infty b_j^2 d^2 H(u) (e_j, e_j)= \tfrac12 B'_1 +\gamma_I \nu X(\tau),
$$
where
$$
B'_r =2\sum b_j^2 {\|e_j\|'}_r^2 =2\sum b_j^2\lambda_j^r \qquad \forall\,r,
$$
and
\begin{equation*}
\begin{split}
X(\tau)&=2p \Re \int \left( |u|^{2q-2} u^2 \sum_j b_j^2 e_j (x)^2\right) dx + \int |u|^{2q} \sum_j b_j^2 e_j(x)^2 dx\\
& \leq CB_0 |u(\tau)|^{2q}_{2q}.
\end{split}
\end{equation*}
Therefore applying Ito's formula we get that 
\begin{equation}\label{1.8}
\begin{split}
dH(u(\tau))&=\Big( (-\gr\lan Au,|u|^{2p}u\ran+\vk\lan Au,u_{xx}\ran
-\gi\nu\gr\int|u|^{2p+2q+2}dx\\
&+\vk\gi\nu\lan|u|^{2q}u,u_{xx}\ran+B_1'+\gi\nu X(\tau)\Big)d\tau+dM(\tau),
\end{split}
\end{equation}
where 
$
dM(\tau)=\sum b_j\lan Au+\gi\nu|u|^{2q}u,e_j\ran\cdot d\bb_j(\tau).
$

Denoting
$
U_q(x)=\frac1{q+1}u^{q+1}$ and  $U_p(x)=\frac1{p+1}u^{p+1},
$
we have 
\begin{equation*} 
\begin{split}
\lan|u|^{2q}u,u_{xx}\ran
\le -\int |u_x|^2|u|^{2q}\,dx=-\|\frac{\p}{\p x}U_q\|^2,
\end{split}
\end{equation*}
and a similar relation holds for $q$ replaced by $p$. Accordingly, 
\begin{equation}\label{1.10}
\begin{split}
dH(u(\tau))\le-\frac12 \Big( \vk \|u\|_2^2+
\gr \|\frac{\p}{\p x}U_p\|^2
 +\vk\gi\nu\|\frac{\p}{\p x}U_q\|^2 \\ 
+\nu\gi\gr\int |u|^{2p+2q+2}dx   - C_\vk\|u\|^2  -2B_1'\Big)\,d\tau+dM(\tau),
\end{split}
\end{equation}
where $C_\vk$ may be chosen independent from $\vk$ if $\gr>0$. Considering relations on
$H(u)^m$, $m\ge1$, which follow from \eqref{1.10} and \eqref{1.8}, using \eqref{1.5} and 
arguing by induction we get that 
\begin{equation}\label{1.11}
\E\left( \sup_{0\le t\le T}H(u(t))^m
+\frac{\vk}2\int_0^TH^{m-1}\|u\|_2^2\,ds\right)\le
H(u_0)^m+C_m(\vk,T,B_1)(1+\|u_0\|^{c_m}),
\end{equation}
\begin{equation}\label{1.12}
\E H(u(t))^m \le C_m(\vk,B_1)(1+ H(u_0)^m+ \|u_0\|^{c_m})\quad \forall\,t>0.
\end{equation}
Estimates \eqref{1.11} in a traditional way (cf. \cite{Hai01b, KS04J, Od06, Sh06}) imply that
 eq.~\eqref{1.1}  is {\it regular in  space $\cH^1$} in the sense that for any $u_0\in\cH^1$
 it has a unique strong solution, satisfying \eqref{1.5}, \eqref{1.11} 
 
 \subsection{Stationary measures.}\label{s1.3}
 The a-priori estimates on solutions of \eqref{1.1} and the Bogolyubov-Krylov argument (e.g.,
 see in \cite{KS}) imply that eq.~\eqref{1.1} has a stationary measure $\mu^\nu$, supported by
 space $\cH^2$. Now assume that 
 \begin{equation}\label{1.160}
b_j\ne0\qquad\forall j.
\end{equation}
Then the approaches, developed in the last decade to study the 2d stochastic Navier-Stokes 
equations, apply to \eqref{1.1} and  allow to prove that under certain restrictions on the  
equation the stationary measure $\mu^\nu$ is unique. In particular this is true if $\gi=0$ (the
easiest case), or if $p\ge q$ and $\gr\ne0$ (see \cite{Od06}), or if $\gr=0$ and $p=1$ (see
\cite{Sh06}). In this case any solution $u(t)$ of \eqref{1.1} with $u(0)=u_0\in\cH^1$ satisfies 
\begin{equation}\label{1.17}
\cD u(t)\strela\mu^\nu\quad\text{as}\quad t\to\infty. 
\end{equation}
This convergence and \eqref{1.3}, \eqref{1.12} imply that 
\begin{equation}\label{1.18}
\int e^{\rho_\vk\|u\|^2}\mu^\nu(du)\le C(\vk,B),
\end{equation}
\begin{equation}\label{1.19}
\int \|u\|_1^{2m}\,\mu^\nu(du)\le C_m(\vk,B_1)\quad\forall\, m.
\end{equation}

\subsection{Multidimensional case.}\label{s2}
In this section we briefly discuss a multidimensional analogy of eq.~\eqref{1.1}:
\begin{equation}\label{2.1}
\begin{split}
\dot u+i\nu^{-1}Au&=\Delta u -\gr  f_p(|u|^{2})u-i\gi  f_q(|u|^{2})u\\
&+\frac{d}{d\tau}\sum_{j=1}^\infty b_j\bb_j(\tau)e_j(x),\qquad 
u=u(\tau,x),\;\; x\in\T^d.
\end{split}
\end{equation}
Here $
Au=-\Delta u+V(x)u$,  $\ V\in C^N(\T^d,\R)$ and $V(x)\ge1$.
The numbers  $\gi,\gr$ satisfy \eqref{1.0}.
Functions $f_p\ge0$ and $f_q\ge0$ are real-valued smooth and 
$$
f_p(t)=t^p\;\;\;\text{for}\;\; t\ge1,\qquad f_q(t)=t^q\;\;\;\text{for}\;\; t\ge1,
$$
where $p,q\ge0$. If $\gr=0$, then  the term $\Delta u$ in the r.h.s. should be modified to $\Delta-u$. 
 By $\{e_k,k\ge1\}$, we denote the usual trigonometric basis of the space 
$L_2(\T^d)$ (formed by all functions $\pi^{-d/2}f_{s_1}(x_1)\dots f_{s_d}(x_d)$,  where each 
$f_s(x)$ is $\sin sx$ or $\cos sx$), 
parameterised by natural numbers.  These are eigen-functions of the Laplacian, 
$-\Delta e_r=\lambda_re_r$. 
We assume that 
\begin{equation}\label{2.3}
B'_{N_1}=2\sum_k \lambda_k^{N_1}b_k^2<\infty, 
\end{equation}
where $N_1=N_1(d)$ is sufficiently large. In this section we denote by $(\cH^r, \|\cdot\|_r)$ the
Sobolev space 
$\ 
\cH^r=H^r(\T^d,\C),
$
regarded as a real Hilbert space, and $\lan\cdot,\cdot\ran$ stands for  the real $L_2$-scalar product. 

Noting that $(f_p(|u|^2)u-|u|^{2p}u)$ and  $(f_q(|u|^2)u-|u|^{2q}u)$  are bounded Lipschitz 
functions with compact support we immediately see that  the a-priori estimates from
 Section~\ref{s1.1}  remain true for solutions of \eqref{2.1}. Accordingly, for any
 $u_0\in\cH^1\cap L_{2q+2}$ eq.~\eqref{1.1} has a solution $u(t,x)$ such that $u(0,x)=u_0$, 
 satisfying \eqref{1.3},  \eqref{1.11},  \eqref{1.12}. 
 
 Now assume that 
\begin{equation}\label{2.4}
p,q<\infty\;\;\;\text{if}\;\;d=1,2,\qquad p,q<\frac{2}{d-2}\;\;\;\text{if}\;\;d\ge3. 
\end{equation}
Applying Ito's formula to the processes $\lan A^mu(\tau),u(\tau) \ran^{n}$,
 $m,n\ge1$, using  \eqref{1.3},  \eqref{1.11},  \eqref{1.12}  and arguing by induction (first in $n$ and next in $m$) we get that 
 \begin{equation}\label{2.5}
 \begin{split}
\E\left(\sup_{0\le\tau\le T}  {\|u(\tau)\|'}_{2m}^{2n}+
\int_0^T{\|u(s)\|'}^2_{2m+1} { \|u(s)\|'}_{2m}^{2n-2}ds\right)\\ 
\le {\|u_0\|'}_{2m}^{2n}+
C(m,n,T)\big(1+\|u_0\|^{c_{m,n}}\big),
\end{split}
\end{equation}
 \begin{equation}\label{2.05}
 \E\,
{\|u(\tau)\|'_{2m}}^{2n} \le  C(m,n)\qquad  \forall\,\tau\ge0, 
\end{equation}
 for each $m$ and $n$, where $C(m,n,T)$  and $C(m,n)$  also depends on $|V|_{C^N}$ and $B_{N_1}$ (see
  \eqref{2.3}), and $N=N(m)$, $N_1=N_1(m)$. 
 
 Relations \eqref{2.5} with $m=m_0\ge1$ in the usual way (cf. \cite{Hai01b,KS04J, Od06, Sh06})  
 imply that eq.~\eqref{2.1} is {\it regular in the space $\cH^{m_0}\cap L_{2q+2}$} in the 
 sense that for any $u_0\in \cH^{m_0}\cap L_{2q+2}$ it has a unique strong solution $u(t,x)$, 
 equal $u_0$ at $t=0$, and satisfying estimates \eqref{1.3}, \eqref{2.5} with $m=m_0$ for 
 any $n$. By the Bogolyubov-Krylov argument this equation has a stationary measure 
 $\mu^\nu$, supported by the space $\cH^{m_0}\cap L_{2q+2}$, and a corresponding 
 stationary solution $u^\nu(\tau)$, $\cD u^\nu(\tau)\equiv\mu^\nu$, also satisfies \eqref{1.3} and
 \eqref{2.05} with $m=m_0$. 
 
If \eqref{1.160} holds and \eqref{2.4} is replaced by a stronger assumption, then a stationary 
measure is unique. If $\gi=0$, the uniqueness readily follows, for example, from the 
abstract theorem in \cite{KS}. In \cite{Sh06} this assertion is proved   if 
\begin{equation}\label{2.6}
\gr=0\;\;\text{and}\;\;q\le1\;\;\text{if}\;\; d=1,\quad q<1\;\;\text{if}\;\;d=2,\quad
q\le2/d\;\;\text{if}\;\; d=3.
\end{equation}
In \cite{Od06} it is established if
\begin{equation}\label{2.7}
p=q,\  \gr,\gi>0\;\;\text{if  $d=1,2$, \ \  and}\;\; p=q<\frac2{d-2},  \  \gr,\gi>0\;\;\text{if}\;\;d\ge3;
\end{equation}
the argument of this work also applies if $p>q$. 
\medskip

Note that when $\gr=0$ or when $p<q$ (i.e., when the nonlinear damping is weaker than
the conservative term), the assumptions \eqref{2.6}, \eqref{2.7}, needed for the uniqueness 
of the stationary measure, are much stronger than the assumptions \eqref{2.4}, needed for
the regularity. This gap does not exist (at least it shrinks a lot) if the random force in
eq.~\eqref{2.1} is not white in time, but is a kick-force. See  in \cite{KS00} the abstract theorem
and its application to the CGL equations. 

\subsection{Spectral properties of  $A_V$: one-dimensional case.}\label{s3.1}

As in Section~\ref{s1.1} we denote $A_V=A=-\p^2/\p x^2+V(x)$, where the potential 
$V(x)\ge0$ belongs to the space $C^N_e$ of $C^N$-smooth even and $2\pi$-periodic 
functions, $N\ge1$. 
 Let $\vp_1,\vp_2,\dots$ be the $L_2$-normalised  complete system of eigenfunctions 
of $A_V$ with the eigenvalues $1\le\lambda_1<\lambda_2<\dots$. Consider the linear 
mapping 
$$
\Psi: \cH\ni u(x)\mapsto v=(v_1,v_2,\dots)\in\C^\infty,
$$
defined by the relation $u(x)=\sum v_k\vp_k(x)$.  In the space of complex sequences $v$
we introduce the norms
$$
|v|^2_{h^m}=\sum_{k\ge1}|v_k|^2\lambda_k^m,\quad m\in\R, 
$$
and denote $h^m=\{v\mid |v|_{h^m}<\infty\}$. Due to the Parseval identity, 
$\Psi:\cH\to h^0$ is a unitary isomorphism. For any $m\in\N$ we have 
$|v|^2_{h^m}=\lan A^m u(x), u(x)\ran$. So the norms $|v|_{h^m}$ and $\|u\|_m$ are 
equivalent for $m=0,\dots,N$.  Since $\Psi^*=\Psi^{-1}$, then the norms are equivalent for 
integer $|m|\le N$. By interpolation they are equivalent for all real  $|m|\le N$. So 
\begin{equation}\label{3.1}
\text{the maps $\quad
\Psi: \cH^m\to h^m,\qquad |m|\le N,\quad$ are isomorphisms.
}
\end{equation}
 Denote $G=\Psi^{-1}:h^m\to\cH^m$. Then
$$
\Psi\circ A\circ G=\diag \{\lambda_k, k\ge1\}=: \hA.
$$
Consider the operator
\begin{equation}\label{3.x}
\cL:= \Psi\circ(-\Delta)\circ G=\Psi\circ(A-V)\circ G=\hA-\Psi\circ V\circ G=:\hA-\cL^0.
\end{equation}
 By \eqref{3.1} $\cL^0=\Psi\circ V\circ G$ defines bounded maps
\begin{equation}\label{3.2}
\cL^0:h^m\to h^m\qquad \forall\, |m|\le N,
\end{equation}
and in  the space  $h^0$ it is selfadjoint. 
\medskip

For any finite $M$ consider the mapping 
$$
\Lambda^M: C_e^N\to \R^M,\qquad V(x)\mapsto (\lambda_1,\dots,\lambda_M). 
$$
Since the eigenvalues $\lambda_j$ are different, the mapping is analytic. As the functions 
$\vp^2_1, \vp_2^2,\dots$ are linearly independent by the classical result of G.~Borg, then for 
any $V\in C_e^N$ the linear mapping 
\begin{equation}\label{4.1}
d\Lambda^M(V):C_e^N\to \R^M\quad \text{is surjective}. 
\end{equation}
In the space $C^N_e$ consider a Gaussian measure $\mu_K$ with a non-degenerate correlation 
operator $K$ (so for  the quadratic  function 
$f(V)=\lan V,\xi\ran_{L_2}\lan V,\eta\ran_{L_2}$  we have 
$\int f(V)\mu_K(dV)=\lan K\xi,\eta\ran$). 
Relation \eqref{4.1} easily implies

\begin{lemma}\label{l4.1}
For any $M\ge1$ the
 measure $\Lambda^M\circ\mu_K$ is absolutely continuous with respect to the Lebesgue 
measure on $\R^M$. 
\end{lemma}

We will call a vector $\Lambda\in\R^\infty$ {\it nonresonant} if for any non-zero integer vector $s$ 
of finite length  we have 
\begin{equation}\label{4.2}
\Lambda\cdot s\ne0.
\end{equation}
A potential $V(x)$ is called nonresonant if its spectrum $\Lambda(V)=(\lambda_1,\lambda_2,\dots)$ is nonresonant. 
The nonresonant potentials are defined in $C_e^N$ by a countable family of open dense relations
\eqref{4.2}. So 
 \begin{equation}\label{4.3}
 \begin{split}
&\text{the nonresonant potentials form a subset of $C_e^N$}\\
&\text{of the second Baire category.}
\end{split}
\end{equation}
Applying Lemma~\ref{l4.1} we also get 
 \begin{equation}\label{4.4}
 \begin{split}
\text{the nonresonant potentials form a subset of $C_e^N$ 
of full $\mu_K$ measure,}
\end{split}
\end{equation}
for any Gaussian measure $\mu_K$ as above. 

The non-resonant vectors $\Lambda$ are important because of the following version of the 
Kronecker-Weyl theorem:

\begin{lemma}\label{l4.2}
Let $f\in C^{n+1}(\T^n)$ for some $n\in\N$. 
 Then for any nonresonant vector $\Lambda$ we have 
$$
\lim_{T\to\infty}\frac1{T}\int_0^T f(q_0+t\Lambda^n)\,dt = (2\pi)^{-n}\int f\,dx,\qquad
\Lambda^n=(\Lambda_1,\dots,\Lambda_n),
$$
uniformly in $q_0\in\T^n$. The rate of convergence depends on $n$, $\Lambda$ 
 and  $|f|_{C^{n+1}}$.
\end{lemma}
\noindent
{\it Proof.} Let us write $f(q)$ as the Fourier series $f(q)=\sum f_se^{is\cdot q}$. Then for 
each non-zero $s$ we have
$\ 
|f_s|\le C_{n+1}|f|_{C^{n+1}}|s|^{-n-1}.
$
So for any $\e>0$ we may find $R=R_\e$ such that 
$
\left| \sum_{|s|>R} f_se^{is\cdot q}\right|\le \frac{\e}2$ for each $q$.  Now it suffices to show that 
 \begin{equation}\label{4.5}
\left| \frac1{T}\int_0^T f_R(q_0+t\Lambda^n)\,dt-f_0\right|\le \frac{\e}2\qquad 
\forall\,T\ge T_\e
\end{equation}
for a suitable $T_\e$, where $f_R(q)=\sum_{|s|\le R}f_se^{is\cdot q}$.  But
$$
\left| \frac1{T} \int_0^Te^{is\cdot(q_0+t\Lambda^n)}\,dt\right|\le\frac2{T|s\cdot \Lambda^n|},
$$
for each nonzero $s$. Therefor the l.h.s. of \eqref{4.5} is
$$
\le \frac2{T}\left(\inf_{|s|\le R}|s\cdot\Lambda^n|\right)^{-1}  \sum|f_s|
\le T^{-1}|f|_{C^0}\, C(R,\Lambda).
$$
Now the assertion follows.
\qed\medskip

\subsection{Spectral properties of  $A_V$: multi-dimensional case.}\label{s3.2}
Now let, as in Section~\ref{s2},  $A=A_V$ be the operator $A=-\Delta+V(x)$, $x\in\T^d$, where 
$1\le V(x)\in C^N(\T^d)$. Let $\{\vp_k(x), k\ge1\}$ be its $L_2$-normalised eigenfunctions and
$\{\lambda_k, k\ge1\}$,  be the corresponding eigenvalues, 
$\ 
1\le\lambda_1\le\lambda_2\le\dots
$.
For any $M\ge1$ denote by $F_M\subset C^N(\T^d)$ the open domain
$$
F_M=\{V\mid \la_1<\la_2<\dots<\la_M\}.
$$
Its complement $F_M^c$ is a real analytic variety in $C^N(\T^d)$ of codimension $\ge2$, so $F_M$ is
connected (see \cite {KK95} and references therein). The functions $\la_1,\dots,\la_M$ are analytic
in $F_M$. Let us fix any non-zero vector $s\in\Z^\infty$ such that $s_l=0$  for $l>M$. The set
$$
Q_s=\{V\in F_M\mid \Lambda(V)\cdot s=0\}
$$
clearly is closed in $F_M$. Since the function $\Lambda(V)\cdot s$ is analytic in $F_M$, then either 
$Q_s=F_M$, or $Q_s$ is nowhere dense in $F_M$. Theorem~1 from \cite{KK95} immediately implies 
that $Q_s\ne F_M$, so \eqref{4.3} also holds true in the case we consider now.

Let $\mu_K$ be a Gaussian measure with a non-degenerate correlation operator, supported by 
the space $C^N(\T^d)$. As   $\Lambda(V)\cdot s$ is a non-trivial  analytic function  on $F_M$ 
  and $F_M^c$ is an analytic variety of positive codimension, 
  then $\mu_K(Q_s)=0$ (e.g., see Theorem~1.6 in \cite{AKSS}).  Since this is true for any $M$ and
  any $s$ as above, then  the assertion \eqref{4.4} also is true. 

\section{Averaging theorem.}\label{s5}
The approach and the results of this section apply both to equations \eqref{1.1} and \eqref{2.1}.
We present it for eq.~\eqref{1.1} and at Subsection~\ref{s5.5} discuss small changes, needed to 
treat \eqref{2.1}. Everywhere below $T$ is an arbitrary fixed positive number.

\subsection{Preliminaries.}\label{s5.1}
In eq. \eqref{1.1} with $u \in\cH^1$ we pass to the $v$-variables, $v=\Psi(u)\in h^1$:
 \begin{equation}\label{5.1}
 \begin{split}
\dot v_k+i\nu^{-1}\lambda_kv_k=P_k(v)\,d\tau +\sum_{j\ge1} B_{kj}\, d\bb_j(\tau),\quad k\ge1.
\end{split}
\end{equation}
Here $B_{kj}=\Psi_{kj}b_j$ (a matrix with real entries, operating on complex vectors), and
 \begin{equation}\label{5.00}
P_k=P_k^1+P_k^2+P_k^3,
\end{equation}
where $P^1,\ P^2$ and $P^3$ are, correspondingly, the linear, dissipative and Hamiltonian 
parts of the perturbation:
\begin{equation*}
 \begin{split}
 P^1(v)=\vk\Psi\circ \frac{\p^2}{\p x^2}u,\;\;\;
 P^2(v)=-\gr \Psi(|u|^{2p}u),\;\;\;
P^3(v)=-i\gi\Psi (|u|^{2q}u),
\end{split}
\end{equation*}
where $u=G(v)$. We will refer to equations \eqref{5.1} as to the {\it $v$-equations.}

For $k\ge1$ let us denote $I_k=I(v_k)=\tfrac12|v_k|^2$ and $\vp_k=\vp(v_k)=$Arg$\,v_k\in S^1$, where
$\vp(0)=0\in S^1$. 
Consider the mappings\begin{equation*}
\Pi_I:h^r\ni v\mapsto I=(I_1,I_2,\dots)\in h^r_{I+},\qquad
\Pi_\vp:h^r\ni v\mapsto \vp=(\vp_1,\vp_2,\dots)\in\T^\infty.
\end{equation*}
Here $h^r_{I+}$ is the positive octant in the space
$$
 h^r_{I}=\{I\mid | I|_{h^r_I}=2\sum_jj^{2r}|I_j|<\infty\}.
$$
We will write 
$$
\Pi_I (\Psi(u))=I(u),\qquad \Pi_\vp(\Psi(u))=\vp(u),\qquad (\Pi_I\times\Pi_\vp)(\Psi(u))=(I\times\vp)(u).
$$
The mapping $I:\cH^r\to h_I^r$ is  2-homogeneous 
continuous, while the mappings $\vp:\cH^r\to \T^\infty$ 
and $(I\times\vp):\cH^r\to h^r_I\times\T^\infty$ are Borel-measurable and discontinuous (the torus
$\T^\infty$ is given the Tikhonov topology and a corresponding distance).

Now let us pass in eq. \eqref{5.1} from the complex variables $v_k$ to the real variables 
$I_k\ge0,\  \vp_k\in S^1$:
 \begin{equation}\label{5.2}
 \begin{split}
dI_k(\tau)
=(v_k\cdot P_k)(v)\,d\tau + Y_k^2\,d\tau  +\sum_l\Psi_{kl}b_l(v_k\cdot d\bb_l),\quad
Y_k=\sqrt{\sum b_l^2\Psi_{kl}^2},
\end{split}
\end{equation}
 and 
 \begin{equation}\label{5.3}
 \begin{split}
d\vp_k(\tau)&=\Big(\nu^{-1}\lambda_k+|v_k|^{-2}(iv_k)\cdot P_k
-|v_k|^{-2}\sum_lb_l(\Psi_{kl}\cdot v_k)(\Psi_{kl}\cdot iv_k)\Big)\,d\tau\\
&+\sum_l |v_k|^{-2}b_l\Psi_{kl}(iv_k\cdot d\bb_l)\\
&=:(\nu^{-1}\lambda_k +G_k(v))\,d\tau +\sum_{l}g_{kl}(v)
\left(\frac{iv_k}{|v_k|}\cdot d\bb_l(\tau)\right).
\end{split}
\end{equation}
 Due to \eqref{3.x}, 
 \eqref{3.2} 
\begin{equation*}
P(v)=\vk\hA v+P^0(v), \qquad P^0:h^r\to h^r\quad \forall\, \tfrac12<r\le N,
\end{equation*}
where  the map $P^0$ is real analytic. The mapping
 $P^0(v)$ and its differential $dP^0(v)$ both have a polynomial growth in $|v|_{h^r}$. Therefore
\begin{equation*}  
|P(v)-P(v^m)|_{r-2-1/3}\le m^{-1/3}Q(|v|_{h^r}),
\end{equation*}
where $Q$ is a polynomial. Here for any 
$v=(v_1,v_2,\dots)\in h^1$ we denote $v^m=(v_1,\dots,v_m)\in\C^m$ and identify it with the 
vector  $(v_1,\dots.v_m,0,\dots)\in h^1$. 

The functions $G_k$ and $g_{kl}$ are singular as $v_k=0$ and satisfy the following estimates:
\begin{equation}\label{5.6}
|G_k(v)\chi_{\{|v_k|>\delta\}}|\le\delta^{-1}Q_k(|v|_r),
\end{equation}
\begin{equation}\label{5.7}
|g_{kj}(v)\chi_{\{|v_k|>\delta\}}|\le\delta^{-1}  j^{-N} Q_{kN}(|v|_r),
\end{equation}
where $Q_k$ and $Q_{kN}$ are polynomials.

For any vector $\theta=(\theta_1,\theta_2,\dots)\in\T^\infty$ we denote by $\Phi_\theta$ the unitary rotation
$$
\Phi_\theta: h^r\to h^r,\qquad v\mapsto v_\theta,\quad\text{where}\quad  v_{\theta\,j}=e^{i\theta_j}v_j\;\;
\forall\,j.
$$
By $\lan F\ran$ etc we denote the averaged functions,
$\ 
\lan F\ran(v)=\int_{\T^\infty}F(\Phi_\theta v)\,d\theta.
$
They are $\vp$-independent, so $\lan F\ran=\lan F\ran(\Pi_I(v))$. The functions 
$\lan P\ran,\lan F\ran,\dots$  also satisfy the estimates above. So 
$$
|\lan(v_k\cdot P_k)\ran(I^m)-\lan(v_k\cdot P_k)\ran(I)|\le  m^{-1/3}C_kQ(|I|_{h^1_I}),
$$
where $Q$ is a polynomial.

Since the dispersion matrix $\{B_{kj}\}$ is non-degenerate, then repeating for equations \eqref{5.1} and \eqref{5.2}
the arguments from Section~7 in \cite{KP08} (also see Section~6.2 in \cite{K10}), we get

\begin{lemma}\label{l5.1} Let $v^\nu(\tau)$ be a solution of \eqref{5.1} and $I^\nu(\tau)=I(v^\nu(\tau))$. Then for 
 any $k\ge1$ the following convergence hold uniformly in $\nu>0$:
\begin{equation}\label{5.8}
\int_0^T\IP\{I^\nu_k(\tau)\le\delta\}\,d\tau\to0\qquad \text{as}\;\; \delta\to0.
\end{equation}
\end{lemma}

(Certainly the rate of the convergence  depends on $k$.)

\subsection{The theorem.}\label{s5.2}
Let us abbreviate 
$$
h^1=h,\quad h^1_I=h_I,\quad C([0,T], h_{I+})=\cH_I,
$$
where $h_{I+}$ is the positive octant $\{I\in h_I\mid I_j\ge0\;\;\forall\,j\}$. Fix any $u_0\in h$. Due to the
estimates \eqref{1.1} and \eqref{1.12} the set of laws $\{\cD(I^\nu(\cdot))\}$, $0<\nu\le1$, is tight in $\cH_I$.
Denote by $Q^0$ any limiting measure as $\nu=\nu_j\to0$, i.e.
\begin{equation*} 
\cD(I^{\nu_j}(\cdot))\strela Q^0
\as \nu_j\to0.
\end{equation*}

Let us consider the averaged drift 
$(\lan(v_k\cdot P_k)\ran(I) +Y_k^2)\,d\tau$ for eq.~\eqref{5.2}. We have 
\begin{equation}\label{5.9}
\lan(v_k\cdot P_k)\ran(v)=\int_{\T^\infty}(e^{i\theta_k}v_k)\cdot P_k(\Phi_\theta v)\,d\theta=v_k\cdot R'_k(v),
\end{equation}
where $R'_k=\int_{\T^\infty}\left( e^{-i\theta_k}P_k(\Phi_\theta v)\right)\,d \theta $
(note that  $\lan(v_k\cdot P_k)\ran$ depends only on $I=\Pi_I(v)$, while $R'_k(v)$ depends on $v$).
The diffusion matrix for \eqref{5.2} is
$
\{ A_{kr}, \  k,r\ge1\},
$ 
where 
\begin{equation*}
 \begin{split}
A_{kr}(v)=\sum_l\left( \Psi_{kr}b_lv_k\right)\cdot\left(\Psi_{rl}b_lv_r\right)
=\sum_lb_l^2(v_k\cdot v_r)\Psi_{kl}\Psi_{rl}.
\end{split}
\end{equation*}
Its average is 
\begin{equation}\label{5.10}
 \begin{split}
\lan A_{kr}\ran(v)&
=\sum_lb_l^2
\int_{\T^\infty}\Re \big(e^{-i(\theta_k-\theta_r)}
v_k \bar v_r\big)\Psi_{kl}\Psi_{rl}\,d\theta\\
&=\delta_{kr}|v_k|^2Y_k^2,\qquad Y_k=\left( \sum_lb_l^2|\Psi_{kl}|^2\right)^{1/2}.
\end{split}
\end{equation}
Due to \eqref{3.1}, 
\begin{equation} \label{5.11}
\sum_k Y_k^2k^{2m}\le C_m B_m \qquad \forall\,m.
\end{equation}

Our first goal is to prove the following averaging theorem:

\begin{theorem}\label{t5.2} The measure $Q^0$ is a solution of the martingale problem
in the space $h_I$ with the drift $(\lan v_k\cdot P_k\ran(I)+Y_k^2)d\tau$ and the diffusion matrix $\lan A_{kr}\ran(I)$. 
That is, $Q^0=\cD(I^0(\cdot))$, where the process $I^0(\tau)$ is a weak (in the sense of stochastic analysis) 
solution of the averaged equations
\begin{equation}\label{5.12}
 \begin{split}
d I_k=\big(\lan(v_k\cdot P_k)\ran(I)+Y_k^2\big)\,d\tau
+\sum_r\big(\sqrt{\lan A\ran}\big)_{kr}(I)\,d\beta_r(\tau),\;\; k\ge1; 
\end{split}
\end{equation}
$ I(0)=I_0=\Pi_I(v_0)$. Moreover, 
\begin{equation}\label{5.13}
\E\sup_{0\le\tau\le T}|I^0(\tau)|^n_{h_I}\le 
C_n(\|u_0\|_1^{2n}+1)\quad\forall\, n,
\end{equation}
\begin{equation}\label{5.14}
\E\int_0^T|I^0(\tau)|_{h^2_I}\,d\tau\le C(\|u_0\|_1^2+1). 
\end{equation}
\end{theorem}
\noindent
{\it Proof.} The crucial step of the proof is to establish the following lemma:

\begin{lemma}\label{l5.3}
Let $\tilde F(v)$ be an analytic function 
on the space $h=h^1$ which extends to an analytic function on $h^{2/3}$ of a polynomial growth. Then 
\begin{equation}\label{5.15}
\mathfrak A^\nu:=\E\max_{0\le\tau\le T}\left|\int_0^\tau\big(\tilde F(I^\nu(s),\vp^\nu(s))
- \lan\tilde F\ran(I^\nu(s))\big)ds\right|\to0 \as \nu\to0.
\end{equation}
\end{lemma}

The lemma is proved  below in Section~\ref{s5.3}, following the argument in \cite{KP08}.
Now we derive from it the theorem.
Let us equip the space $\cH_I$ with the Borel sigma-algebra $\cF$, the natural filtration of sigma-algebras 
$\{\cF_\tau, 0\le\tau\le T\}$ and the probability $Q^0$. The fact that the processes
 $I^\nu_k(\tau)-\int_0^\tau \big( (v_k^\nu \cdot P_k)(v^\nu)  +Y_k^2\big) \,ds\ $ are martingales, the convergence 
 $\cD(I^{\nu_j}(\cdot))\strela Q^0$ and  Lemma~\ref{l5.3} with $\tilde F=F_k$ imply that the processes 
 $\ Z_k(\tau)=I_k(\tau)-\int_0^\tau\big(\lan(v_k\cdot P_k)\ran(I(s) +Y_k^2\big)\,ds,\,  k\ge1, $  
  are $Q^0$-martingales, cf. Section~6 of \cite{KP08}. 
 
 Similar to \eqref{5.15} we find that 
 $$
 \E\max_{0\le\tau\le T}\left|\int_0^\tau\big(\tilde F(I^\nu(s),\vp^\nu(s))
- \lan\tilde F\ran(I^\nu(s))\big)ds\right|^4
\to0 \as \nu\to0.
 $$
Then using the same arguments as before, we see that the processes
$Z_k(\tau)Z_j(\tau)-\int_0^\tau\lan A_{kj}\ran(I(s))\,ds$ 
also are $Q^0$-martingales. That is, $Q^0$ is a solution of the martingale problem with the drift 
 $\lan F_k\ran +Y_k^2$
and the diffusion $\lan A\ran$. Hence, $Q^0$ is a law of a weak solution of eq.~\eqref{5.12}. See \cite{Yor74}.

Estimates \eqref{5.13}, \eqref{5.14} follow from \eqref{1.11} and the basic properties of the weak convergence 
since $\|u\|_m^2\sim |v|_m^2 =   \Pi_I(v)|_{h_I^m}$.
\qed

\subsection{Proof of Lemma \ref{l5.3}.}\label{s5.3}
Fix any $m\ge1$ and denote by $I^{\nu, m}, \vp^{\nu,m}$ etc the vectors, formed by the first $m$ components 
of the infinite vectors $I^\nu, \vp^\nu$, etc. Below $R$ denotes a suitable function of $\nu$ such that 
$R(\nu)\to\infty$ as $\nu\to0$, but
\begin{equation}\label{5.150}
R^N\nu\to0 \as \nu\to0,\qquad \forall\, N.
\end{equation}
Denote by $\Omega_R=\Omega_R^\nu$ the event
$$
\Omega_R=\{\sup_{0\le\tau\le T}|v^\nu(\tau)|_{h_1}\ge R\}.
$$
Then  $\IP(\Omega_R)\le\vk_\infty(R)$ uniformly in $\nu$ (see Notations). We denote 
$$
\IP_{\OR}(Q)=\IP(\OR^c\cap Q),\qquad \E_{\OR}(f)=\E \big(f \chi_{\OR^c}\big).
$$
Since for $|v|_{h^1}\le R$ we have $|v-v^m|_{h^{2/3}}\le C(R)m^{-1/3}$ and since $\tilde F$ is Lipschitz on 
$h^{2/3}$, then
$$
\frak A^\nu\le \vk_\infty(R)+C_k(R)m^{-2/3}+\E_{\OR}
\max_{0\le\tau\le T}
\left| 
\int_0^\tau\big(\tilde F(I^{\nu,m},   \vp^{\nu,m})-   \lan\tilde F\ran^m(I^{\nu,m})\big) \, ds\right|.
$$
Here $\lan\tilde F\ran^m$ stands for averaging of the function
$\T^m\ni I^m\mapsto \tilde F(I^m,0,\dots)$. 
So it remains to estimate for any $m$ and $R$  an analogy 
 $\frak A^\nu_{m,R}$ 
 of the quantity $\frak A^\nu$ for the finite-dimensional process 
$I^{\nu,m}(\tau)$ on the event $\OR$ (where its norm is $\le R$),
\begin{equation*}
\frak A^\nu_{m,R}=
\E_{\OR}
\max_{0\le\tau\le T}
\left| \int_0^\tau\big(\tilde F(I^{\nu,m},   \vp^{\nu,m})-   \lan\tilde F\ran^m(I^{\nu,m})\big) \, ds\right|.
\end{equation*}

Consider a partition of $[0,T]$  by the points
$$
\tau_j=\tau_0+jL,\quad 0\le j\le K,
$$
where $\tau_{K}$ is the last point $\tau_j$ in $[0,T)$.  The diameter $L$ of the partition is
$$
L=\sqrt\nu, 
$$
and the non-random phase $\tau_0\in(0,L]$ will be chosen later.  Denoting
\begin{equation}\label{5.17}
\eta_l=
\int_{\tau_l}^{\tau_{l+1}}\big(\tilde F(I^{\nu,m},   \vp^{\nu,m})-   \lan\tilde F\ran^m(I^{\nu,m})\big) \, ds,\quad
0\le l\le K-1,
\end{equation}
we see that 
\begin{equation}\label{5.170}
\frak A^\nu_{m,R} \le LC(R)+\E_{\OR}\sum_{l=0}^{K-1}|\eta_l|,
\end{equation}
so it remains to estimate $\E_{\OR}\sum |\eta_l|$. 
We have
\begin{equation*}
 \begin{split}
|\eta_l|  & \le \left|
\int_{\tau_l}^{\tau_{l+1}}\big(\tilde F(I^{\nu,m}(s),   \vp^{\nu,m}(s))-   \tilde F (I^{\nu,m} (\tau_l), \vp^{\nu,m}
(\tau_l)+\nu^{-1}\Lambda^m(s-\tau_l
 )) \big) ds\right|\\
&+ \left|
\int_{\tau_l}^{\tau_{l+1}}\big(
 \tilde F (I^{\nu,m} (\tau_l), \vp^{\nu,m}
(\tau_l+\nu^{-1}\Lambda^m(s-\tau_l))-
 \lan\tilde F\ran^m(I^{\nu,m} (\tau_l) )
 \big) \, ds\right|\\
& + \left|
\int_{\tau_l}^{\tau_{l+1}}\big(
 \lan\tilde F\ran^m(I^{\nu,m} (\tau_l) - \lan\tilde F\ran^m(I^{\nu,m} (s) 
 \big) \, ds\right| =:\Upsilon^1_l+\Upsilon^2_l+\Upsilon^3_l.
\end{split}
\end{equation*}
To estimate the quantities $\Upsilon^j_l$ we first optimise the choice of the phase $\tau_0$. Consider the events 
$\cE_l, 1\le l\le K  $,  
\begin{equation}\label{5.151}
\text{$\cE_l=\{I_k^\nu(\tau_l)\le\gamma\}, \quad$  where 
$\gamma\ge\nu^a,\quad  a=1/10.$}
\end{equation}
By Lemma~\ref{5.1} and the Fubini theorem 
we can choose $\tau_0\in[0,L)$ in such a way that 
\begin{equation*} 
K^{-1}\sum_{l=0}^{K-1}\IP(\cE_l)=\vk(\gamma^{-1}; R,m).
\end{equation*}
For any $l$ consider the event 
$$
Q_l=\{\sup_{\tau_l\le\tau\le\tau_{l+1}}|I^\nu(\tau)-I^\nu(\tau_l)|_{h_I}\ge P_1(R)L^{1/3}\},
$$
where $P_1(R)$ is a suitable polynomial. It is not hard to verify (cf. \cite{KP08}) that 
$\IP(Q_l)\le \vk_\infty(L^{-1})$.
Setting 
$$
\cF_l=\cE_l\cup Q_l   
$$
we have that 
\begin{equation*} 
\frac1{K}\sum_{l=0}^{K-1}\IP(\cF_l)\le \vk(\gamma^{-1}; R,m)+\vk(\nu^{-1/2};m)=:\tilde\vk. 
\end{equation*}
Accordingly,
$$
\frac1{K}\sum_{l=0}^{K-1}\left|(\E-\E_{\cF_l})\Upsilon_l^j\right| \le \frac{P(R)}{K}\sum_{l=0}^{K-1}\IP(\cF_l)\le
P(R)\tilde\vk:=\tilde\vk_1,\quad j=1,2,3.
$$

Similar, 
since for $\omega\not\in\OR$ the integrand in \eqref{5.17} is $\le Q(R)$, then 
\begin{equation}\label{5.21}
\frac1{K}\sum_l\left| \E_{\OR}\eta_l-\E_{\OR\cup\cF_l}\eta_l\right| \le \tilde\vk Q(R).
\end{equation}
If $\omega\notin\OR\cup \cF_l$, then for $\tau\in[\tau_l,\tau_{l+1}]$ we have that  $I^\nu_k(\tau_l)\ge\gamma-
P_1(R)L^{1/3}\ge\tfrac12\gamma$,  if $\nu$ is small. This relation and \eqref{5.3}, \eqref{5.6}, \eqref{5.7} imply that 
\begin{equation*}
 \begin{split}
\IP_{\OR\cup \cF_l} \{ |\vp^{\nu, m}(s)-(\vp^{\nu,m}(\tau_l)+\nu^{-1}\Lambda^m(s-\tau_l)|
\ge \nu^a\;\;\text{for some}&\;\; s\in[\tau_l,\tau_{l+1}]
\}\\
&\le \vk^\infty(\nu^{-1};R,m). 
\end{split}
\end{equation*}
Accordingly, 
\begin{equation}\label{5.001}
 \left(\sum_l  
\E_{\OR\cup\cF_l}  \Upsilon^1_l\right) \le C\nu^a+\vk^\infty(\nu^{-1};R,m).
\end{equation}
It is clear that 
\begin{equation}\label{5.002}
 \left(\sum_l  
\E_{\OR\cup\cF_l}  
 \Upsilon^3_l\right) \le P(R)L^{1/3}=P(R)\nu^{1/6}.
\end{equation}
So it remains to estimate the expectation of $\sum\Upsilon^2_l$. For any $\omega\notin\OR\cup\cF_l$ abbreviate 
$$
F(\psi)=\tilde F(I^{\nu,m}(t_l),\vp^{\nu,m}(t_l)+\psi),\qquad \psi\in\T^m,
$$
where  in the r.h.s. $\psi$ is identified with  the vector $(\psi,0,\dots)\in\T^\infty$. 
We can write $\Upsilon^2_l$ as 
$$
\Upsilon^2_l=\left|\int_{\tau_l}^{\tau_{l+1}}F(\nu^{-1}\Lambda^m(s-\tau_l))\,ds-\lan F\ran
\right|=L
\left|  \frac{\nu}{L} \int_{0}^{\nu^{-1}L} F(\Lambda^m t)\,dt-\lan F\ran
\right|.
$$
Since the function $F(\psi)$ is analytic and the vector $\Lambda$ is non-resonant,  then by Lemma~\ref{l4.1}
$\ 
\Upsilon^2_l\le L\vk(\nu^{-1}L;m,R,\gamma,\Lambda).
$
Therefore
\begin{equation}\label{5.003}
 \left(\sum_l  
\E_{\OR\cup\cF_l}
\Upsilon^2_l\right) \le \vk(\nu^{-1/2};m,R,\gamma,\Lambda). 
\end{equation}

Now \eqref{5.170},  \eqref{5.21} and  \eqref{5.001}- \eqref{5.003} imply that 
\begin{equation*}
 \begin{split}
\frak A^\nu\le \vk_\infty(R)+C(R)m^{-1/3}+\vk(\nu^{-a};R,m)+\vk(\gamma^{-1};R,m)\\
+C\nu^a+P(R)\nu^{1/6}+\vk(\nu^{-1/2};m, R, \gamma,\Lambda).
\end{split}
\end{equation*}
Choosing first $R$ large, then $m$ large and next $\gamma$ small and
$\nu$ small in such a way that \eqref{5.150} and \eqref{5.151} hold, 
 we make the r.h.s. arbitrarily small. This proves the lemma.

\subsection{Joint distribution of actions and angles. }\label{s5.4}

Denote $\tilde\mu_s^\nu=\cD(I^\nu(s),\vp^\nu(s))=(I\times \vp)\circ\cD(u^\nu(s))$, 
where   $u^\nu(s),\,  0\le s\le T$,  is a solution of \eqref{1.1} and $(I^\nu,\vp^\nu)$ is a solution 
of the system \eqref{5.2}, \eqref{5.3}.   For any $f\in L_1(0,T),\  f\ge0$,  such that $\int f=1$, set 
$\tilde\mu^\nu(f)=\int_0^Tf(s)\tilde\mu^\nu_s\,ds$.  Also let us denote 
$m^0(f)=\int_0^Tf(s)\cD(I^0(s))\,ds$; this is a measure on $h_{I+}$.

\begin{theorem}\label{t5.4} 
For any $f$ as above,
\begin{equation}\label{5.22}
\tilde\mu^{\nu_j}
(f)\strela m^0(f)\times d\vp \as  \nu_j\to0.
\end{equation}
\end{theorem}
\noindent
{\it Proof.}  For a piecewise constant function $f$ the convergence follows from Theorem~\ref{t5.2} and
Lemma~\ref{l5.3} since  by the lemma, for any $0\le T_1<T_2\le T$, 
 the integral 
$ \int_{T_1}^{T_2}\tilde F(I^\nu(s),\vp^\nu(s))\,ds$
is close to 
$
 \int_{T_1}^{T_2} \lan \tilde F  \ran  (I^\nu(s))\,ds,
$
and by the theorem the integral $ \int_{T_1}^{T_2} \lan \tilde F  \ran  (I^\nu(s))\,ds$ is close to
$ \int_{T_1}^{T_2} \lan \tilde F  \ran  (I^0 (s))\,ds =  \int_{\T^\infty} \int_{T_1}^{T_2}\tilde F(I^0 (s),\psi)\,ds\,d\psi$
(we are applying the lemma and the theorem 
on segments $[0,T_1]$ and $[0,T_2]$).

To get the convergence for a general function $f$ we approximate it by 
piecewise constant functions. See Section~\ref{s5} of \cite{K10} for details. \qed.

\subsection{Multidimensional case. }\label{s5.5}

Let \eqref{5.1} be not eq.~\eqref{1.1}, but eq.~\eqref{2.1}, written in the $v$-variables. Now we should 
consider \eqref{5.1} as an equation in a space $h^r,\ r>d/2$. The maps $P^1:h^r\to h^r$ and 
$P^2:h^r\to h^r$ are smooth and the differentials $d^mP^1(v):h^r\times\dots\times h^r\to h^r$ are 
poly-linear mappings such that their norms are bounded by polynomials of $|v|_{h^r}$.  This allows to 
apply to eq.~\eqref{5.1} the methods of \cite{KP08}
\footnote{In was assumed in \cite{KP08} that the relevant maps and vector-fields are analytic. This
analyticity was imposed only for simplicity. Sufficiently high smoothness and polynomial estimates on
the corresponding high order differentials are sufficient for all construction of \cite{KP08}.}
in the same way as in Sections~\ref{5.2}-\ref{5.3} and establish validity of Theorems~\ref{t5.2} and 
\ref{t5.4}.

\section{Effective equations and uniqueness of  limit.   }\label{s7}
Let \eqref{5.1} be eq. \eqref{1.1} or eq. \eqref{2.1}, written in the $v$-variables, and \eqref{5.12} -- 
the corresponding averaged equation. Accordingly, by $h$ we denote either the space $h^1$ as in 
Section~\ref{s1}, or the space $h^r$, $r>d/2$, as in Section~\ref{s2}. For simplicity we assume that $p$
and $q$ in \eqref{2.1} are integers. If they are not, then in the calculations below the nonlinearities 
$|u|^{2p}u$ and  $|u|^{2q}u$ should be modified by Lipschitz terms which cause no extra difficulties. 

\subsection{Effective equations.}\label{s7.1}
Let us write the averaged drift 
 $\lan v_k\cdot P_k\ran$ and the averaged diffusion $\lan A_{kr}\ran$
 in the form \eqref{5.9}
and \eqref{5.10}, respectively. Using \eqref{5.00} we write the term $R'(v)$  in \eqref{5.9} 
as 
$$
R'_k(v)=\sum_{m=1}^3\int e^{-i\theta_k}P^m_k(\Phi_\theta v)\,d\theta=:
\sum_{m=1}^3R^m_k(v),\qquad k\ge1.
$$
By  \eqref{3.x} and 
\eqref{3.2},
\begin{equation}\label{7.1}
\begin{split}
R^1(v&)=\vk\int \Phi_{-\theta}\Psi\left(\frac{\p^2}{\p x^2}G(\Phi_\theta v)\right)\,d\theta\\
&=-\vk\int \Phi_{-\theta}  \hA\Phi_\theta v\,d\theta+\vk\int \Phi_{-\theta} \cL^0(\Phi_\theta v)\,d\theta\\
&=-\vk\hA+\vk R^0(v),\qquad R^0(v)=\int \Phi_{-\theta} \cL^0(\Phi_\theta v)\,d\theta,
\end{split}
\end{equation}
since $\hat A$ commutes with the rotations $\Phi_\theta$. 
The operator $R^0$ is bounded and selfadjoint in $h^0$.
For any $v$ we have 
\begin{equation}\label{7.01}
\lan R^1(v),v\ran=\vk \int \lan\Delta G\Phi_\theta v,G\Phi_\theta v\ran\,d\theta\le 
-C\vk|v|^2_{h^1},
\end{equation}
since $\|G\Phi_\theta v\|_1\sim |\Phi_\theta v|_{h^1}=|v|_{h^1}$. Writing in \eqref{7.1}
 $\cL^0(v)=\Psi\circ V\circ G(v)$ as
$$
\cL^0(v)=\vk\nabla(h^2\circ G)(v),\qquad h^2(u)=\frac12\int V(x)|u(x)|^2dx,
$$
we have $R^0(v)=\nabla\lan h^2\circ G\ran(v)$. Since 
\begin{equation*}
\begin{split}
\lan h^2\circ G\ran(v)&=\frac12\sum_{j,l}\int_{\T^\infty}\lan V(x)e^{i\theta_j}v_j \vp_j(x),
 e^{i\theta_l}v_l \vp_l(x)\ran\,d\theta\\
& =\frac12 \sum_l |v_l|^2M_l,\qquad M_l=\lan V\vp_l,\vp_l\ran,
\end{split}
\end{equation*}
then 
$
R^0=\diag\{M_l, \  l\ge1\}.
$
Accordingly,
\begin{equation}\label{7.02}
R^1=\vk\diag\{-\lambda_l+M_l,\ l\ge1\}>0,\qquad M_l=\lan V\vp_l,\vp_l\ran
\end{equation}

The term $R^2$ is defined as an integral with the integrand 
$$
\Phi_{-\theta}P^2\Phi_\theta(v)=
-\gr\Phi_{-\theta}\Psi(|u|^{2p}u) \mid_{u= G\circ\Phi_\theta v}
=: F_\theta(v).
$$
Writing $f^p(|u|^2)u$
\footnote{if $d=1$ and $p$ is an integer, than $f^p(|u|^2)=|u|^{2p}$.}
as $\nabla h^p(u)$, where $h^p(u)=\int F^p(|u|^2)\,dx$,  $(F^p)'=\tfrac12f^p$, and 
denoting $G\circ \Phi_\theta=L_\theta$, we have 
$$
F_\theta(v)=-\gr L_\theta^*
\nabla h^p(u)\mid_{u=L_\theta(v)}=-\gr\nabla(h^p \circ L_\theta(v)).
$$
So 
\begin{equation}\label{7.2}
R^2(v)=-\gr \nabla_v\left( \int_{\T^\infty} (h^p\circ G)(\Phi_\theta v)\,d\theta
\right) =\gr\nabla_v \lan h^p\circ G\ran.
\end{equation}

Similar 
$\ 
R^3(v)=-i\gi\nabla_v\lan h^q\circ G\ran
$ 
(since the operator $G\circ\Phi_\theta$ is complex-linear). As  $\lan h^q\circ G\ran$ is a function solely 
of the actions $(I_1,I_2,\dots)$, then $\nabla_{v_k}\lan h^q\circ G\ran \in\C$ is a vector, 
real-proportional to $v_k$. Therefore $v_k\cdot R^3_k(v)=0$ for each $k$. That is,
\begin{equation}\label{7.2x}
\lan(v_k\cdot P_k)\ran(v)=v_k\cdot R^1_k(v)+v_k\cdot R^2_k(v),
\end{equation}
where $R^1$ and $R^2$ are defined by \eqref{7.02} and \eqref{7.2}. Now we set
$$
R(v)=R^1(v)+R^2(v)
$$
and consider the following system of stochastic equations:
\begin{equation}\label{7.3}
dv_k(\tau)=R_k(v)\,d\tau + Y_k\,d\bb_k,\qquad k\ge1.
\end{equation}
Equations \eqref{7.3} are called the {\it system of effective equations}. 
\medskip

\begin{example}\label{e7.1} ($p=1$).
Now $h^1(u)=\frac14\int |u|^4dx$. So 
$$
h^1\circ G(v)=\frac14\int \left| \sum_kv_k\vp_k(x)\right|^4dx=
\frac14\sum_{k_1,k_2,k_3,k_4} v_{k_1}v_{k_2}\bar v_{k_3}\bar v_{k_4}  \int\vp_{k_1}
\vp_{k_2}\vp_{k_3} \vp_{k_4}\,dx.
$$
Since
$$
\lan  v_{k_1}v_{k_2}\bar v_{k_3}\bar v_{k_4} \ran=
\begin{cases}
|v_{k_1}|^2|v_{k_2}|^2
&\quad \text{if $k_1=k_3, k_2=k_4$ or $k_1=k_4, k_2=k_3$},
\\
0&\quad \text{otherwise},
\end{cases}
$$
then
\begin{equation}\label{7.03}
\lan h^1\circ G(v)\ran=\frac12\sum_{k_1\neq k_2}|v_{k_1}|^2 |v_{k_2}|^2 L'_{k_1k_2}+
\frac14\sum_k|v_k|^4L'_{kk},
\end{equation}
where
$\ 
L'_{k_1k_2}=\int \vp_{k_1}^2\vp_{k_2}^2\,dx.
$
So that 
\begin{equation*}
\begin{split}
R^2_k(v)&=-\gr\nabla_{v_k}\lan h^1\circ G\ran(v)=
-\gr v_k\left( |v_k|^2L'_{kk}  +  2\sum_{l\ne k}   |v_l|^2L_{kl}'    \right)    \\
&=
-\gr v_k\sum_l|v_l|^2L_{kl}.
\end{split}
\end{equation*}
Here $L_{kk}=L'_{kk}$ and   $L_{kl}= 2L'_{kl}$  if $k\ne l$. 
So the system of effective equations becomes
\begin{equation}\label{7.6}
dv_k=-v_k\big(\vk (\lambda_k-M_k)+\gr\sum_l|v_l|^2L_{kl}\big)\,d\tau
+Y_k\,d\bb_k,\qquad k\ge1.
\end{equation}
\qed
\end{example}

If $v(\tau)=\{v_k(\tau),\ k\ge1\}$ satisfies \eqref{7.3}, then for $I_k=I(v_k(\tau))$ we have 
\begin{equation}\label{7.4}
dI_k(\tau)=
v_k\cdot R_k(v)\,d\tau + Y_k^2\,d\tau+  Y_k v_k\cdot d\bb_k,\qquad k\ge1.
\end{equation}
By \eqref{7.2x} the 
 drift in this system equals $(\lan v_k\cdot P_k \ran(I)+Y_k^2)d\tau$, while the diffusion matrix is 
$\ \delta_{kr}|v_k|^2Y_k^2=\lan A_{kr}\ran$. So system \eqref{7.4} has the same set of weak (=
 martingale) solutions as \eqref{5.12}, see \cite{Yor74}. We have got

\begin{proposition}\label{p7.1}
Let $v(\tau)$ be a weak solution of \eqref{7.3} such that $v(0)=v_0$ and 
\begin{equation}\label{7.5}
\E\sup_{0\le\tau\le T}|v(\tau)|_h^{2n}\le C|v_0|_h^{2n}+C(n,T),\quad \forall\,n.
\end{equation}
Then $\Pi_I(v(\tau))$ is a weak solution of the system \eqref{5.12}, satisfying \eqref{5.13} and such that
$I(0)=I_0$. 
\end{proposition}
That is, the solutions of eq.~\eqref{5.12} which can be obtained as limits (when $\nu\to0$) of
actions $I^\nu(u(\tau))$ of solutions for \eqref{1.1}  (or \eqref{2.1})
are those which can be covered by ``regular"  solutions of \eqref{7.3}.
\medskip

The `right' inverse statement  to Proposition~\ref{p7.1}
is given by the following

\begin{proposition}\label{t7.2}
Let $I^0(\tau)$ be a weak solution of the averaged equations \eqref{5.12}, constructed in 
Theorem~\ref{t5.2}. Then there exists a weak solution $v^0(\tau)$ of \eqref{7.3} such that 
$v(0)=v_0$, satisfying \eqref{7.5}, and such that $\cD\big(\Pi_I(v^0(\cdot))\big)=\cD(I^0(\cdot))$.
\end{proposition}

For a proof we refer to Section~3 of \cite{K10}, where the assertion is established in a similar
but more complicated situation.
\smallskip

System \eqref{7.3} is invariant under rotations $\Phi_\theta$:

\begin{proposition}\label{p7.2}
Let $v(\tau)$ be a weak solution of \eqref{7.3}, satisfying \eqref{7.5}. Then, for any $\theta\in\T^\infty$,
$\Phi_\theta v(\tau)$ is a weak solution of \eqref{7.3}, satisfying \eqref{7.5}.
\end{proposition}
\noindent
{\it Proof.} Applying $\Phi_\theta$ to \eqref{7.3} we get that 
$$
d(\Phi_\theta v)=\Phi_\theta R(v)\,d\tau + \Phi_\theta Y\,d\bb(\tau),\qquad Y=\diag \{Y_k\}.
$$
The vector fields $R^1(v)$ and $R^2(v)$ both are obtained by averaging and have the form 
$R^j(v)=\int\Phi_{-\theta}F^j(\Phi_\theta v)\,d\theta$.  So they commute with the rotations, as well as their sum 
$R(v)$, and we have
$$
d(\Phi_\theta v)=R(\Phi_\theta v)\,d\theta + Y\,d(\Phi_\theta \bb(\tau)).
$$
Since $\cD\Phi_\theta \bb(\tau)=\cD \bb(\tau)$, then the assertion follows. \qed 
\medskip

\subsection{The uniqueness.}\label{s7.2}
Let $v^1(\tau)$ and $v^2(\tau)$ be solutions of the effective system \eqref{7.3}.  Denoting
$v=v^1-v^2$, we have that 
$$
\frac12 \frac{d}{d\tau}|v(\tau)|^2_{h^0} \le -  \vk |v|^2_{h^1}+
\lan R^2(v^1)-R^2(v^2),v^1-v^2\ran.
$$
Consider the last term, denoting $v^j_\theta=\Phi_\theta v^j,\ u^j_\theta=G(v^j_\theta)$.
Since $R^2(v)$ is  an integral over $\T^\infty$ 
 with the integrand
$-\gamma_R\Phi_{-\theta}\Psi(|u_\theta|^{2p}u_\theta)$, where ${u_\theta=G(\Phi_\theta(v))}$, 
then 
\begin{equation*}
\begin{split}
\lan R^2(v^1)-R^2(v^2),v^1-v^2\ran &= -\gr\int\big \lan\Psi(|u^1_\theta|^{2p}u^1_\theta - 
|u^2_\theta|^{2p}u^2_\theta ), \Phi_\theta v^1-\Phi_\theta v^2\big\ran\,d\theta\\
&=-\gr\int\lan ( |u^1_\theta|^{2p}u^1_\theta -  |u^2_\theta|^{2p}u^2_\theta, u^1_\theta-u^2_\theta
\big\ran\,d\theta.
\end{split}
\end{equation*}
The  integrand  in the r.h.s. is non-negative. So 
\begin{equation}\label{7.8}
\frac12  \frac{d}{d\tau}|v|^2_{h^0}\le -C  \vk |v|^2_{h^1},
\end{equation}
(i.e., the effective system  \eqref{7.3} is strongly monotone).
Therefore  a strong solution of the system \eqref{7.3} is unique. By the Yamada-Watanabe 
argument (see \cite{KaSh}) a weak solution also  is unique. 
We have got

\begin{theorem}\label{t7.3}
Let $I^\nu(\tau)=I(u^\nu(\tau))$, where $u^\nu(\tau)$ is a solution of eq.~\eqref{1.1} or of 
eq.~\eqref{2.1} and $u^\nu(0)=u_0$. Then 
$$
\cD(I^\nu(\cdot))\strela Q^0 \as \nu\to0 
$$
in the space $\cH_I$, where $Q^0$ is a weak solution of \eqref{5.12}, satisfying \eqref{5.13}, 
\eqref{5.14}. There exists a unique weak solution $v(\tau)$ of the effective equations \eqref{7.3}, 
satisfying \eqref{7.5},  such that $v(0)=\Psi(u_0)$ and $\cD(\Pi_I(v(\cdot))=Q^0$. 
\end{theorem}

\section{Stationary solutions.}\label{s6}
\subsection{Averaging.}\label{s6.1}
Again, 
let \eqref{5.1} be eq. \eqref{1.1} or eq. \eqref{2.1}, written in the $v$-variables, and \eqref{5.12} be
the corresponding averaged equation. Accordingly, by $h$ we denote either the space $h^1$ as in 
Section~\ref{s1}, or the space $h^r$, $r>d/2$, as in Section~\ref{s2}.  
Assume that the corresponding 
$u$-equation is regular in the space $\cH^r$ (e.g., $d=1$ or the assumptions, given at the end of the
Section~\ref{s2} are fulfilled), and that it has a unique stationary measure $\mu^\nu$ (see 
Sections~\ref{s1.3}, \ref{s2}).

Let ${u'}^\nu(\tau)$ be a stationary in time solution of equation \eqref{1.1}, $ \cD({u'}^\nu(\tau))\equiv \mu^\nu$.
 By estimates in Section~\ref{s1} the set of laws
 $\cD({I'}^\nu(\cdot))$,  where ${I'}^\nu=I({u'}^\nu(\tau))$, is compact in $h_I$. Let $Q'$ be any limiting measure 
 as $\nu_j\to0$. Clearly it is stationary in $\tau$. The same argument that was used to prove 
 Theorem~\ref{t5.2} (cf. \cite{KP08}) imply that $Q'$ is a stationary solution of the averaged equation:

 \begin{proposition}\label{t6.1}
 The measure $Q'$ is the law of a process $I'(\tau),\, 0\le\tau\le T$, which is a stationary weak solution of the 
 averaged equation \eqref{5.12}. It satisfies estimates \eqref{5.13}, \eqref{5.14}, and the stationary measure 
 $\pi=\cD(I'(0))$ meets estimates $\eqref{1.17}$, $\eqref{1.18}$ with $\nu=0$. 
 \end{proposition}
 
 The measures $(I\times \vp)\circ\mu^\nu=\cD({I'}^\nu(s),{\vp'}^\nu(s))$  satisfies \eqref{5.22}
 for the same reason as in Section~\ref{s5.4}. Since the measure $\mu^\nu$ is independent from $s$,
 then now
 \begin{equation}\label{6.1}
 \cD({I'}^\nu(s),{\vp'}^\nu(s))\strela \pi\times d\vp\as \nu_j\to0.
\end{equation} 
 In the stationary case relation \eqref{5.8} implies that 
 \begin{equation}\label{6.2}
 \IP\{{I'}^\nu_k(\tau)<\delta\}\to 0 \as \delta\to0,
\end{equation} 
uniformly in $\nu$. In particular,
\begin{equation}\label{6.02}
 \pi\{I\mid I_k=0\}=0\qquad\forall\,k.
\end{equation}

 \subsection{Lifting to effective equations.}\label{s6.2}
 To study the limiting measure $\pi$ further we lift it to a stationary measure of the effective system \eqref{7.3}.
 We start with
 
 \begin{lemma}
 Assume \eqref{1.160}. Then the system \eqref{7.3} has a unique stationary measure $m$.
 \end{lemma}
\noindent
{\it Proof.} Relation \eqref{1.160} implies that $Y_k\ne0$ for all $k$. That is, the noise in the effective 
equations is non-degenerate. Moreover, the coefficients $Y_k$ satisfy \eqref{5.11}. Since solutions of
\eqref{7.3} satisfy estimates \eqref{7.5} and since any two solutions converge by \eqref{7.6}, then the 
assertion follows.  E.g., see \cite{KS}, Section~3. \qed
\medskip

Let $v(\tau)$ be a stationary solution of \eqref{7.3}, $\cD(v(\tau))\equiv m$.  By Proposition~\ref{7.2}, 
$\Phi_\theta(v(\tau))$ also is a (weak) stationary solution. So 
$\cD(\Phi_\theta v(\tau))=\Phi_\theta\circ m$ is  a stationary measure for \eqref{7.3}. Since it is unique, then 
$$
\Phi_\theta\circ m=m\qquad \forall\,\theta\in\T^\infty.
$$
Accordingly, $\,\Pi_\vp\circ m$ is a rotation-invariant measure on $\T^\infty$, i.e.
 $\,\Pi_\vp\circ m=d\vp$.  This implies that 
  in the $(I,\vp)$-variables the measure $m$ has the form 
\begin{equation}\label{6.5}
 dm=m_I(dI)\times d\vp
\end{equation}

Proposition  \ref{6.1} applies for any time-interval $[0,T]$. So, replacing the sequence $\nu_j\to0$ by
a suitable subsequence $\nu_{j'}\to0$ we construct a stationary process $I'(\tau)$, $\tau\ge0$, 
such that ${I'}^{\,\nu_{j'}}(\tau)$ converges to $I'(\tau)$ in distribution on any finite time-interval. Using
Theorem~\ref{7.2} and discussion at the end of Section~\ref{7.1} we construct a solution $v'(\tau)$ of
\eqref{7.3} such that $\cD(\Pi_I(v'(\tau))\equiv\pi$. Since $\cD(v'(\tau))\strela m$ as $\tau\to\infty$, then
\begin{equation}\label{6.6}
\pi=\Pi_I\circ m.
\end{equation} 
That is, the measure $\pi$ is independent from the sequence $\nu_j$. We have got

\begin{theorem}\label{t6.3}
If \eqref{1.160} holds, then $I\circ\mu^\nu\strela \pi=\Pi_I\circ m$, where $m$ is the unique 
stationary measure of the effective system.
\end{theorem}

In view of \eqref{6.1}, \eqref{6.5} and \eqref{6.6},
$$
(I\times \vp)\circ\mu^\nu \strela (\Pi_I\times\Pi_\vp)\circ m\as \nu\to0.
$$
Denote $h_+=\{v\in h\mid v_j\ne0\ \forall\,j\}$.  By \eqref{6.2} and \eqref{6.02}
 $\ (\Psi\circ\mu^\nu)(h_+)=1$ and $m(h_+)=1$. So the convergence above implies that 
 
 \begin{theorem}\label{t6.4}
 If \eqref{1.160} holds, then $\mu^\nu\strela G\circ m$ as $\nu\to0$.
 \end{theorem}
 
 \begin{example}  \label{e6.3} (Hamiltonian perturbations.)
 If in \eqref{2.1} $\gr=0$, i.e. if the nonlinear term of the perturbation is Hamiltonian, then the 
 effective system is the linear equation
 $$
 dv(\tau)=R^1(v)\,d\tau+Y\,d\bb,
 $$
 where $R^1$ is defined in \eqref{7.02} and $Y=\diag\{Y_k,\  k\ge1\}$. Let $v(0)=0$. Then $v(\tau)$ is 
 the diagonal complex Gaussian process
 $$
 v(\tau)=\int_0^\tau e^{(\tau-s)R^1}Y\,d\bb(s),\qquad R^1=\vk(\hA -R^0).
 $$
 So the stationary measure for the effective system, $\cD v(\infty)$, is a direct sum of independent 
 complex Gaussian measures with zero mean and the dispersions $\vk^{-1}Y_k^2/(\lambda_k-M_k)$,
 $k\ge1$. 
 
 The fact that a Hamiltonian nonlinearity produces no effect in the first order averaging (i.e. for the slow time
 $\tau \lesssim 1$) is well known in the theory of weak turbulence.  
  To produce a non-trivial effect, the 
 Hamiltonian term  $ -i\gi f_q(|u|^2)u$ should be scaled by the additional factor $\nu^{-1/2}$, and for the weak 
 turbulence theory to apply to calculate this effect we should send the size of the $x$-torus to
 infinity when $\nu\to0$, see \cite{Naz}. 
\end{example}

\begin{example}  \label{e6.4} ($p=1$, continuation). 
If $p=1$,  then the effective equations become
\begin{equation}\label{6.8}
dv_k=-v_k\big( \vk (\lambda_k-M_k)+\gr\sum_l|v_l|^2 L_{kl} \big)\,d\tau +Y_k\,d\bb_k.
\end{equation} 
Assume that the random force in \eqref{1.1} (or in \eqref{2.1}) is small and is mostly 
concentrated at a frequency $j_*$. That is,
$$
b_{j_*}=\e<1,\quad 0<b_l\ll\e\quad\text{if}\quad l\ne j_*.
$$
Then the numbers $Y_k$ are of order $\e$ and are concentrated close to $j_*$, i.e., 
$$
Y_{j_*}\sim\e,\qquad Y_l\le\e C_N|l-j_*|^{-N}\quad\forall\, l,N.
$$
So if $v(\tau)$ is a stationary solution of the effective equations and $E_k=\tfrac12\E |v_k(\tau)|^2$, 
then 
$$
E_{j_*}\sim\e^2 \lambda_{j_*}^{-1},
\qquad E_l\le\e^2 C_N   \lambda_{j_*}^{-1}  |l-j_*|^{-N}\quad\forall\, l,N.
$$
That is, the systems \eqref{1.1} and \eqref{2.1} exhibit no inverse or direct cascade of energy. 
For other polynomial systems  \eqref{1.1} and \eqref{2.1}   situation is the same. Certainly this is not 
surprising since by imposing the non-resonance condition we removed from the system resonances, 
responsible for the two energy cascades.
\end{example}

\section{Equations with non-viscous damping.}\label{s8}

Following Debussche-Odasso \cite{DO05} we now discuss equations \eqref{1.1}  with non-viscous 
damping, i.e. with $\vk=0$ but with $\gr>0$ and $p=0$ (Debussche-Odasso considered 
the case $p=0, q=1$):
\begin{equation}\label{8.1}
\begin{split}
\dot u+i\nu^{-1}(-u_{xx}+V(x)u)=-\gr u-i\gi|u|^{2q}u
+\frac{d}{d \tau} \sum b_j \bb_j( \tau)e_j(x), \\
u(x)\equiv u(x+2\pi)\equiv -u(-x);
\end{split}
\end{equation} 
\begin{equation}\label{8.0}
u(0)=u_0. 
\end{equation} 
Estimates \eqref{1.5},  \eqref{1.11} and  \eqref{1.12} are valid with $\vk=0$. Jointly with an analogy of
estimate  \eqref{2.5} with $\vk=0, m=1$ they imply that for $u_0\in\cH^2$ the set of actions 
$I^\nu(\tau)=I(u^\nu(\tau))$ of solutions for  \eqref{8.1} ,  \eqref{8.0} is tight in $\cH_I$. As in 
Section~\ref{s5}, any limiting measure $Q^0=\lim \cD(I^{\nu_j}(\cdot))$ is a law of a weak solution 
$I^0(\tau)$ of the averaged equations  \eqref{5.12}${}_{\vk=0}$ with $I(0)=I_0=I(u_0)$. 
Constructions of Section~\ref{s7} remain true, so $I^0(\tau)$ may be lifted to a weak solution 
$v^0(\tau)$ of the effective equations  \eqref{7.3}${}_{\vk=0,\,p=0}$. Now $R^1=0$ and, repeating 
constructions of  Example~\ref{e7.1} we  see that 
$\ 
R^2_k(v)=-\gr  v_k.
$
So the effective equations become the linear system
\begin{equation}\label{8.3}
dv_k(\tau)=-\gr  v_k\,d\tau + Y_k\,d\bb_k.
\end{equation} 
This system has a unique solution $v(\tau)$ such that $v(0)=v_0=\Psi(u_0)$. So
$$
\lim_{\nu\to0} \cD(I^\nu(\cdot))=\cD\Pi_I(v(\cdot)).
$$

Due to results of \cite{DO05},  eq. \eqref{8.1} has a unique stationary measure $\mu^\nu$. 
Repeating arguments from Example~\ref{e6.3}, we see that when $\nu\to0$, the measures 
$\Psi\circ\mu^\nu$ 
converge to the unique stationary measure of eq.~\eqref{8.3} which is 
$$
m=\cD\int_{-\infty}^0 \diag\{e^{-s\gr }Y_k\}\,d\bb_k(s).
$$
This is a direct sum of independent complex gaussian measures with zero mean and the 
dispersion $Y_k^2/\gr , \, k\ge1$.  So every solution $u(\tau)$ of \eqref{8.1} satisfies the 
Gaussian limit 
$$
\lim_{\nu\to0} \lim_{\tau\to\infty}  \cD u(\tau)= G\circ m. 
$$
If we replace in \eqref{8.1} the linear damping by the nonlinear term $-\gr|u|^2u$, then the 
effective system \eqref{8.3} should be replaced by the nonlinear system \eqref{6.8} with 
$\lambda_k=M_k=0$. In this case the limiting measure is non-Gaussian.

\bibliography{meas}

\providecommand{\bysame}{\leavevmode\hbox to3em{\hrulefill}\thinspace}
\providecommand{\MR}{\relax\ifhmode\unskip\space\fi MR }
\providecommand{\MRhref}[2]{%
  \href{http://www.ams.org/mathscinet-getitem?mr=#1}{#2}
}
\providecommand{\href}[2]{#2}
\begin{thebibliography}{AKSS07}

\bibitem[AKN89]{AKN}
V.~Arnold, V.~V. Kozlov, and A.~I. Neistadt, \emph{Mathematical {A}spects of
  {C}lassical and {C}elestial {M}echanics}, Springer, Berlin, 1989.

\bibitem[AKSS07]{AKSS}
A.~Agrachev, S.~Kuksin, A.~Sarychev, and A.~Shirikyan, \emph{On
  finite-dimensional projections of distributions for solutions of randomly
  forced{ PDE}s}, Ann. I. H.Poincar\`e - PR \textbf{43} (2007), 399--415.

\bibitem[DO05]{DO05}
A.~Debussche and C.~Odasso, \emph{Ergodicity for the weakly damped stochastic
  non-linear {S}hr\"odinger equations}, J. Evolut. Eq. \textbf{5} (2005),
  317--356.

\bibitem[FW98]{FW98}
M.~Freidlin and A.~Wentzell, \emph{Random {P}erturbations of {D}ynamical
  {S}ystems}, 2nd ed., Springer-Verlag, New York, 1998.

\bibitem[Hai02]{Hai01b}
M.~Hairer, \emph{Exponential mixing properties of stochastic {PDE}'s through
  asymptotic coupling}, Probab. Theory Relat. Fields \textbf{124} (2002),
  345--380.

\bibitem[Kha68]{Khas68}
R.~Khasminski, \emph{On the avaraging principle for {I}to stochastic
  differential equations}, Kybernetika \textbf{4} (1968), 260--279, (in
  Russian).

\bibitem[KK95]{KK95}
T.~Kappeler and S.~Kuksin, \emph{Strong nonresonance of {S}chr\"odinger
  operators and an averaging theorem}, Physica D \textbf{86} (1995), 349--362.

\bibitem[KP08]{KP08}
S.~B. Kuksin and A.~L. Piatnitski, \emph{Khasminskii - {W}hitham averaging for
  randomly perturbed {K}d{V} equation}, J. Math. Pures Appl. \textbf{89}
  (2008), 400--428.

\bibitem[KS91]{KaSh}
I.~Karatzas and S.~Shreve, \emph{Brownian {M}otion and {S}tochastic
  {C}alculus}, 2nd ed., Springer-Verlag, Berlin, 1991.

\bibitem[KS00]{KS00}
S.~B. Kuksin and A.~Shirikyan, \emph{Stochastic dissipative {PDE}s and {G}ibbs
  measures}, Comm. Math. Phys. \textbf{213} (2000), 291--330.

\bibitem[KS04]{KS04J}
\bysame, \emph{Randomly forced {CGL} equation: stationary measures and the
  inviscid limit}, J. Phys. A: Math. Gen. \textbf{37} (2004), 1--18.

\bibitem[KS10]{KS}
\bysame, \emph{Mathematics of 2d {S}tatistical {H}ydrodynamics}, preprint of a
  book (2010), www.math.polytechnique.fr/$\sim$kuksin/books.html.

\bibitem[Kuk10]{K10}
S.~B. Kuksin, \emph{Damped-driven {KdV} and effective equations for long-time
  behaviour of its solutions}, GAFA \textbf{20} (2010), 1431--1463.

\bibitem[LM88]{LochM}
P.~Lochak and C.~Meunier, \emph{Multiphase {A}veraging for {C}lassical
  {S}ystems}, Springer-Verlag, New York--Berlin--Heidelberg, 1988.

\bibitem[Naz11]{Naz}
S.~Nazarenko, \emph{Wave {T}urbulence}, Springer, Berlin, 2011.

\bibitem[Oda06]{Od06}
C.~Odasso, \emph{Ergodicity for the stochastic complex {G}inzburg-{L}andau
  equations}, Ann. Inst. H.~Poincar\'e - PR \textbf{42} (2006), 417--454.

\bibitem[Shi06]{Sh06}
A.~Shirikyan, \emph{Ergodicity for a class of {M}arkov processes and
  applications to randomly forced {PDE}'s. {II}}, DCDS-A \textbf{6} (2006),
  911--926.

\bibitem[Yor74]{Yor74}
M.~Yor, \emph{Existence et unicit\'e de diffusion \`a valeurs dans un espace de
  {H}ilbert}, Ann. Inst. Henri Poincar\'e Sec.~B, \textbf{10} (1974), 55--88.

\end{thebibliography}
\bibliographystyle{amsalpha}
\end{document}